\documentclass[galaxies,review,accept,oneauthor,pdftex]{mdpi} 

\firstpage{1} 
\pubvolume{8}
\issuenum{0}
\articlenumber{6}
\pubyear{2020}
\copyrightyear{2020}
\history{Received: 13 November 2019; Revised: 30 December 2019; Accepted: 3 January 2020; Published: 8 January 2020}
\usepackage{aas_macros}
\usepackage{color}

\Title{Applications of Stellar Population Synthesis in the Distant Universe}

\Author{Elizabeth R. Stanway $^{1}$\orcidA{0000-0002-8770-809X}}%, Firstname Lastname $^{1,\ddagger}$ and Firstname Lastname $^{2,}$*}

\AuthorNames{Elizabeth R. Stanway}

\address{%
$^{1}$ \quad Department of Physics, University of Warwick, Gibbet Hill Road, Coventry, CV4 7AL, UK; e.r.stanway@warwick.ac.uk}
\abstract{Comparison with artificial galaxy models is essential for translating the incomplete and low signal-to-noise data we can obtain on astrophysical stellar populations to physical interpretations which describe their composition, physical properties, histories and internal conditions. In particular, this is true for distant galaxies, whose unresolved light embeds clues to their formation and evolution as well as their impact on their wider environs. Stellar population synthesis models are now used as the foundation of analysis at all redshifts, but are not without their problems. Here we review the use of stellar population synthesis models, with a focus on applications in the distant Universe.}


\begin{document}
%%%%%%%%%%%%%%%%%%%%%%%%%%%%%%%%%%%%%%%%%%

%%%%%%%%%%%%%%%%%%%%%%%%%%%%%%%%%%%%%%%%%%

\section{Introduction}
The basic premise of stellar population synthesis is not a new one \citep[e.g.][]{1968ApJ...151..547T,1975MmSAI..46....3T,1976ApJ...203...52T}. If an unresolved stellar cluster or galaxy is made of stars, then its light must represent the integrated total light emitted by those stars. By combining models or empirical data for stars of different masses and ages, the properties of populations with known properties can be predicted. The prediction that best fits the unresolved light of the distant population is deemed to best describe its stellar population. Thorough discussions of this stellar population synthesis method, its strengths and key limitations, exist in the literature \citep[e.g.][and references therein]{2012IAUS..284....2L,2013ARA&A..51..393C,2017PASA...34...58E}. 

Evolutionary stellar population synthesis models are amongst the most widely used tools for understanding the properties of distant galaxies and their evolution through cosmic time.  They inform the conversion factors used to convert diagnostic fluxes to star formation rates, they are used to calibrate stellar population age and metallicity diagnostics in the optical and ultraviolet spectrum, they are used for fitting of spectral energy distributions (SEDs) to determine redshifts, stellar masses and ages from limited photometric data, and they are used to populate the dark matter haloes of cosmological simulations with realistic stellar populations which can evolve over time. However different models of stellar populations at the same age and metallicity have been demonstrated to vary in their predictions for the colours and spectral properties of the integrated light  \citep[see e.g.][]{2018MNRAS.476.4459D,2010A&A...515A.101C,2010ApJ...712..833C,2017PASA...34...58E}. This results from the wide range of physics which must be considered in these models, in areas which include the internal structure evolution of stars,  the effects of their binary interactions, rotation, atmospheres and abundance patterns and many others. Depending on the prescription in use, the same stellar evolution tracks or isochrones may be matched to different atmosphere models, or vice versa, producing different interpretations of any given observable. At the same time, the way stellar populations are combined or processed by dust or gas before observations are simulated can contribute further ambiguity, which inevitably requires further astronomical observations, usually over a wider wavelength range or with a higher signal-to-noise, to break.

In the photon-starved regime occupied by observations of the earliest galaxies to radiate through the Universe, this presents a problem: often it simply is not possible to find an independent diagnostic which will distinguish between models, and obtainable signal-to-noise may be limited by instrumental and background noise rather than by the source. In this situation, the interpretation offered by a single population synthesis model is often accepted without close scrutiny, with a deep faith placed on the calibration of those models in the relatively local Universe. In the past, for observations of galaxies at $z<2$, this faith has been well founded. The stellar populations which can be resolved in detail and probed in the Milky Way and the Local Group are typical of galaxies at these redshifts in both age and metallicity. However as the observational frontier is pushed back to ever higher redshifts, it is clear that the typical star formation densities and environments are very different, as are the stellar metallicities \citep[][and references therein]{2014ARA&A..52..415M}. This has prompted a reevaluation of the stellar population synthesis models in use at these redshifts, and their calibration in hitherto unconsidered regimes \citep{2009MNRAS.400.1019E}. 

This review discusses both our current state of understanding of galaxies in the distant Universe, and how that understanding is informed by the stellar population synthesis models we use. Key examples and uncertainties are highlighted, and a holistic approach, in which all possible diagnostic indicators of a stellar population are considered, is advocated.

%%%%%%%%%%%%%%%%%%%%%%%%%%%%%%%%%%%%%%%%%%
\section{Galaxies in the Distant Universe}

\subsection{The First Stars}

The definition of the distant Universe varies, but in the context of galaxy formation and evolution it applies primarily to the interval between the formation of the first stars (`cosmic dawn') which ended the cosmic Dark Ages at around $z\sim7-15$, and the coincident peaks of the volume-averaged star formation rate density history and AGN activity at $z\sim2$ (`cosmic noon'). Over this $\sim$2.5\,Gyr interval, the first galaxies were not only born, but also evolved significantly \citep[e.g.][]{2019arXiv191004179P,2015ApJ...803...34B}, such that galaxy populations at $z>5$ are typically younger, more highly ionizing, less massive and less abundant than their lower redshift counterparts \citep[see][and references therein]{2011ARA&A..49..525S,2016ARA&A..54..761S}.

At the start of this epoch, the earliest stars, known as Population III stars, formed from gas which was metal-free, except for trace amounts of heavy elements which resulted from Big Bang nucleosynthesis. 
Since the primordial gas clouds were unable to cool through metal line emission, their Jeans masses were significantly higher than those typical in today's molecular clouds. As a result, these first stellar populations were likely dominated by a population of very massive, very short lived objects, which ended their lives in massive explosions \citep{2004ARA&A..42...79B}. These created and distributed metals which enriched the gas from which later generations of stars formed \citep[see e.g.][]{2018MNRAS.475.4396J}. The population III epoch of star formation in any given halo was likely brief, although  metal-free stars may have continued to form in isolated and underdense regions as late as $z\sim6$ \citep{2018MNRAS.479.4544M}.
The short time scales involved from massive star birth to metal return to the halo, together with the dense neutral gas in which they were embedded, suggests that observing such sources may never be possible. While population III-dominated galaxy candidates have been identified in observations \citep[e.g.][]{2015ApJ...808..139S}, none have proven robust to later investigation (either theoretical \citep{2017MNRAS.469..448B} or follow-up observations \citep{2018PASJ...70S..15S,2019MNRAS.482.2422S}).

\subsection{The Lyman Break Technique}

The majority of galaxy candidates in the distant Universe have been identified as star-forming sources with a bright rest-frame ultraviolet stellar continuum. Against this, passage through HI clouds in the intergalactic medium (IGM) imposes a forest of absorption lines, primarily between source rest-frame wavelengths of 912\,\AA\ (the ionization limit of hydrogen) and 1216\,\AA\ (the Lyman-$\alpha$ transition of hydrogen) and known as the Lyman-$\alpha$ forest.  The cumulative effect of these lines is to introduce near-complete scattering of light below the 912\,\AA\ Lyman limit and a second break in the wavelength-averaged flux of the Lyman-forest region relative to that longwards of Lyman-$\alpha$. The strength of this second break depends on the number density and column density of H\,I clouds along the line of sight, and so is redshift dependent. By about z$\sim$3.5, half of the flux in the Lyman-forest region is scattered away. By $z\sim5$, the fraction of flux lost exceeds 90 percent \citep[e.g.][]{2002AJ....123.2183S}. The term "Lyman break", while originally applying to the effect of the 912\,\AA\ Lyman limit, is now used colloquially to refer to both of these step features in the spectrum.

The significance of these features lies primarily in the fact that for sources at redshifts $z>3$ the Lyman break is redshifted into the observed-frame optical. As a result, sources in the distant Universe show distinctive optical colours which result from the abrupt spectral breaks falling within or between adjacent filters. By selecting a set of two or three filters which bracket the Lyman break, rare distant galaxy candidates can be identified from amidst the large numbers of faint, nearby sources and prioritised for spectroscopic follow up \citep{1990ApJ...357L...9G,1995AJ....110.2519S,1996ApJ...462L..17S}. Such galaxies are known as Lyman break galaxies (LBGs) or dropouts (for their tendency to vanish from, or drop out of, the bluest images). This method, or variants upon it, have now been applied to select galaxy candidates from $z\sim1.5$ to $z\sim10$ by their photometric colours \citep{2004ApJ...604..534S,2003MNRAS.342..439S,2011MNRAS.417..717W,2014ApJ...793..115B,2019ApJ...880...25B}.

Inevitably, any photometric technique is vulnerable to questions over both contamination and completeness. Assessing the latter requires both a thorough understanding of observational biases and limitations for any given survey, and extrapolations regarding the intrinsic properties (e.g. in size, colour etc) of the population \citep[e.g.][]{2003ApJ...593..640B}. The former issue, that of contamination, is more challenging still to reliably assess. The original $z\sim3-4$ LBG selection proved remarkably clean, since very few other astronomical objects mirror the colour and other properties of galaxies at those redshifts \citep{1995AJ....110.2519S}. The same is not true at all redshifts, with $z\sim5-6$ samples, for example, prone to contamination from both cool Galactic stars and intermediate redshift galaxies with strong line emission in certain bands \citep[e.g. ][]{2008MNRAS.385..493S,2017ApJ...836..239V}. Each time the Lyman break technique is applied in a different redshift or luminosity regime, the potential contaminant population differs, and because the parameter spaces being explored are novel, it is seldom well understood before the observations take place.

Thus colour selection is often combined with other observations, usually either an extremely deep bluewards `veto' filter, in which any detection rules out a very high redshift identification, or a redwards filter which determines that the sources are intrinsically blue (as expected for star forming galaxies) rather than red (as might be the case for a cool star or dusty local contaminant). With sufficient filters considered, this becomes indistinguishable from a simple photometric redshift approach. Additionally, the colour photometry may be combined with tools such as gravitational lensing to boost the signal to noise of the faint sources, or with a narrowband image which identifies flux excesses corresponding to Lyman-$\alpha$ in emission at the target redshift. Nonetheless, the only way of converting a distant galaxy candidate to a known distant galaxy remains a spectroscopic redshift from either the continuum break or an absorption or emission feature in the galaxy spectrum.

\subsection{High redshift galaxy spectroscopy}

Spectroscopy of distant galaxies is challenging. The sources are faint, typically with observed frame optical detections fainter than 25th magnitude, and extending down to 30th magnitude and fainter in the most sensitive deep fields \citep[e.g.][]{2004ApJ...607..704S,2004MNRAS.355..374B,2014MNRAS.443.2831C}. Spreading this light into a spectrum inevitably results in a low photon count rate per pixel, a regime which is dominated by uncertainties not on the photon count, but rather on the instrumental and atmospheric background. In addition, by $z\sim5$ the Lyman break is observed at 7300\,\AA, and at $z\sim8$ it lies beyond 1\,$\mu$m. Since most information on distant sources can only be observed longwards of this feature, the spectral information required lies at the red end of the optical and extends into the infrared. At these wavelengths, the sky background forms a dense forest of bright emission lines, generated by OH- radicals in the Earth's atmosphere and varying in strength not only from site to site but from observation to observation.  Observing at high spectral resolution allows researchers to effectively work between the sky lines, but risks spreading the faint source flux still thinner and cannot exclude the possibility of source features coinciding in wavelength with sky emission. Observing at lower spectral resolution means fighting to distinguish spectral features against sky subtraction residuals. These problems are exacerbated when observing in the near-infrared, where each photon carries too little energy to excite photons across the silicon band gap, and CCD detector sensitivities have traditionally lagged behind those typical in the optical.

Nonetheless, long (typically $\sim$5 hour or longer) observations with large (usually 8 or 10m diameter) telescopes have made such observations almost routine in recent years \citep[e.g.][]{2019MNRAS.488..706H,2019MNRAS.487.2038C,2016ApJ...822...29F,2014A&A...563A..58T}. There are now large samples of galaxies at $z\sim2-6$ with spectroscopic confirmation of their redshifts, and a growing sample with more detailed spectral information. For a redshift, a low signal to noise detection of either the continuum break or the Lyman-$\alpha$ emission line seen in about half the population is usually sufficient. To learn anything more about the stellar population and its properties, multiple emission or absorption lines must be detected and calibrated. 

At $z>4$ the dominant emission feature in the ultraviolet spectrum is often Lyman-$\alpha$, however the resonant scattering of this line and the effective suppression of its blue wing by line-of-sight absorption makes interpretation of the line strength and profile problematic. At slightly lower redshifts ($z=1-3$) where the blue component can be observed, the separation of peaks in the Lyman-$\alpha$ line is a good diagnostic of high ionizing photon escape fractions \citep{2015A&A...578A...7V}. Longwards of Lyman-$\alpha$ in the rest-frame ultraviolet, relatively strong emission lines in [O\,III] 1665\AA, C\,III] 1909\AA, and occasionally He\,II 1640\AA\ or C\,IV 1550\AA, provide diagnostics of the ionizing spectrum in the form of nebular emission with a low velocity width \citep[e.g.][]{2010ApJ...719.1168E,2015MNRAS.450.1846S,2017A&A...608A...4M}, while much broader absorption components of C\,IV 1550\AA\ and emission components of He\,II 1604\AA\ can provide a direct insight into the presence of massive hot stars driving winds, such as Wolf-Rayet stars \citep{2016ARA&A..54..761S,2012MNRAS.419..479E,2003ApJ...588...65S}. For bright sources, or those with particular deep spectroscopy, full spectrum fitting may also be possible, constraining the blanketing of the continuum by iron absorption lines. Sufficient signal to noise to perform such an analysis is rare at $z>2$ but has been obtained by dedicated surveys \citep[e.g.][]{2016ApJ...826..159S,2019MNRAS.487.2038C}. 

Key to doing this is the ability to multiplex spectroscopy, such that many objects can be observed simultaneously in a single long integration or set of integrations, rather than dividing the available telescope time between them. Multiplexing spectroscopy of galaxies requires either a large field of view, or a high source density. The latter is often achieved by combining observations of galaxy candidates across a wide redshift range or selected through different methods into a single program. The former requires technological innovation to develop field correctors and large format CCD detectors, and is still a work in progress for most large telescopes. Nonetheless, deep, long-integration programmes multiplexing to obtain tens of spectra at a time, if not hundreds, are now routine \citep{2010MNRAS.409.1155D,2019MNRAS.488..706H,2019MNRAS.487.2038C}. 

One of the larger steps forward for observations of stellar populations in recent times has resulted from the advent of multiplexed \textit{near-infrared} spectrographs, particularly MOSFIRE on Keck. These have been used most effectively to observe the rest-frame optical spectrum of galaxies close to Cosmic Noon (i.e. $z\sim1-3$) \citep{2019arXiv190700013S,2017ApJ...836..164S,2015ApJ...801...88S,2014ApJ...795..165S}. At this epoch the source density and redshifted wavelength range are optimised to efficiently extract the same diagnostic nebular line emission in the distant galaxy population that is used to evaluate stellar populations nearby.

\subsection{Distant Galaxies as probes of star formation}

The methods discussed to this point are crucial for identifying distant galaxy candidates and for determining their redshifts, while evaluating any contamination to the sample.  Given a survey of Lyman break galaxies, together with a completeness analysis, the number density of sources per unit volume and their luminosity function can be determined. In combination, these provide a volume-averaged ultraviolet luminosity density at a given redshift. In practice, for redshifts $z>4$, this is almost always performed on photometrically-selected LBG samples rather than spectroscopically-confirmed subsamples, in the expectation that observational biases can be corrected on a statistical rather than case-by-case level \citep[e.g.][]{2015ApJ...803...34B}.

\begin{figure}[!th]
\centering
\includegraphics[width=12 cm]{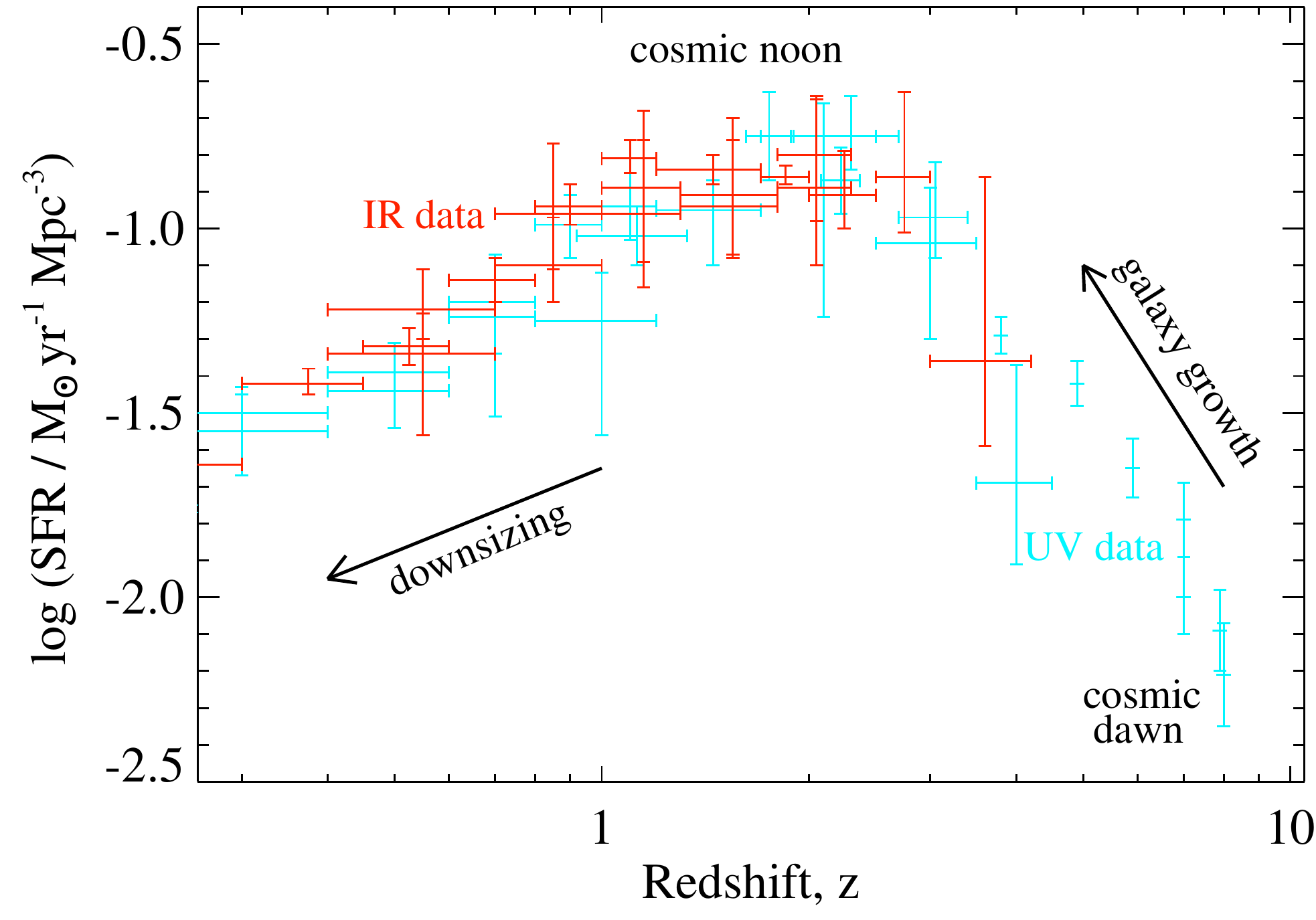}
\caption{The cosmic, volume-averaged star formation rate density history as a function of redshift. Data points show the compilation of \citet[][see also references therein]{2014ARA&A..52..415M}, with infrared-derived and dust-corrected ultraviolet-derived star formation rates distinguished by colour. The volume-averaged flux measured by each data set has been calibrated by \citet{2014ARA&A..52..415M} against stellar population synthesis models. UV-derived rates are also subject to large corrections for assumed dust extinction, which remains uncertain at the highest redshifts. The rise in volumetric star formation rate from cosmic dawn to the peak at cosmic noon, followed by a decline as star formation downsizes, is clearly visible. \label{fig:sfh}}
\end{figure}

The result of decades of work in this field is an overview of cosmic history in which the density of ultraviolet photon production has been traced as a function of time. Since ultraviolet photons are emitted from the atmosphere of recently-formed stars, with ages $<$100\,Myr, this is interpreted as a cosmic star formation rate density history as shown in Figure \ref{fig:sfh} \citep[see][and references therein]{2014ARA&A..52..415M}.  This started low in the very distant Universe, rising rapidly towards $z\sim8$ as dark matter haloes converted their gas into stars. Between $z\sim8$ and $z\sim2$ the rate of change became much slower, accretion and energetic feedback from both supernovae and AGN activity nearly balancing one another to produce a steadily rising plateau in star formation rate density. At $z\sim2$, the cosmic star formation rate density peaks, and at the same time both galaxy merger activity and quasar activity are also peaking. The result is that the gas supply in large galaxies becomes exhausted or is driven out by strong feedback processes, leaving only smaller galaxies able to continue forming stars. This downsizing process \citep{1996AJ....112..839C} is marked by a sharp decline in the volume-averaged cosmic star formation rate density, which drops towards the low levels seen in the local Universe.

Despite this clear and convincing picture, it is often forgotten that interpreting high redshift source densities as a star formation history introduces a wide range of calibration and interpretational assumptions and involves an implicit use of stellar population synthesis models. Uncertainties arise in the statistical corrections for completeness and contamination already discussed, but also in assessing the mean impact of dust extinction on the ultraviolet continuum emission of these sources. Synthesis models come into play when the corrected flux densities are converted to star formation rates \citep[see][]{1998ARA&A..36..189K,2012ARA&A..50..531K}. Historically, this has involved the use of a multiplicative constant; \textit{this} many ultraviolet (1500\,\AA) photons imply \textit{this} star formation rate. However as our understanding of stellar populations becomes more nuanced, it is not clear that such a reductive approach can be supported \citep[e.g.][]{2019MNRAS.tmp.2490W,2017MNRAS.470..489G}.

\subsection{The Critical Importance of Ionizing Flux}

The extreme youth and low metallicities of galaxies in the distant Universe mean that they are highly efficient producers of photons below 912\,\AA, with energies sufficiently high to ionize atomic hydrogen. As Figure \ref{fig:seds} illustrates, the flux shortwards of this Lyman limit drops off rapidly with increasing age or metallicity in stellar population synthesis models. This flux can rarely if ever be observed since it is absorbed and reprocessed by intervening hydrogen. However it is this ionizing energy output from distant sources which arguably has the largest impact  both on the observed properties of 
galaxies and on the wider universe.

Crucially some fraction of this flux must escape the early galaxies hosting star formation in order to reionize the universe at $z\sim6-9$. This transition between the highly neutral intergalactic medium of the Dark Ages and the highly ionized intergalactic medium that has persisted since just after Cosmic Dawn is almost certainly driven by star formation, rather than AGN activity. Indeed much of the observational effort which aims to characterise the luminosity function and star formation rate densities of the Lyman break galaxy population has been geared towards constraining this process \citep[e.g.][]{2004MNRAS.355..374B,2015ApJ...803...34B}. Theoretical models constructed to explore reionization need not only a galaxy density, but also to assume both an ionizing photon production rate for a given unit of star formation and the fraction of photons produced which escape into the intergalactic medium. While a handful of rare objects have been found in which the escape fraction is high \citep[e.g.][]{2016MNRAS.461.3683I}, the typical value for distant galaxies appears to be $<10$ percent \citep[although note that discussion of the biases in these measurements is ongoing, ][]{2013ApJ...765..118W,2014A&A...569A..78V,2017MNRAS.469.3252P,2017ApJ...841L..27R,2015MNRAS.451.2544P}. As a result, it appears that the photon production rate must be high \citep[e.g.][]{2016MNRAS.458L...6W}. A complication in interpretation arises as the flux can only be observed longwards of the Lyman break (typically at around 1500\,\AA) and the photon production shortwards of 912\,\AA\ must be inferred from this. The inference is heavily dependent on the assumed stellar population, and in particular on the modelling of the massive star population, as  Figure \ref{fig:seds} demonstrates.

While constraining reionization is an important goal in observational cosmology, galaxy evolution studies also aim to understand the physical conditions in star forming regions within the distant galaxies themselves. Since the ultraviolet and optical spectra and SEDs of young galaxies are heavily modified by ionized gas and dust, the strength of the ionizing photon flux becomes a critical parameter for understanding their stellar populations. Indeed, both photometric and spectroscopic studies have shown that galaxies with strong recombination line emission and other evidence for a hard ionizing spectrum, such as strong [O\,III], He\,II or C\,III] emission lines and their ratios, are ubiquitous in the distant Universe \citep{2014ApJ...784...58S,2016ApJ...820...73H,2017ApJ...836..164S,2017MNRAS.464..469S,2019MNRAS.489.2355D}. These line ratios (in the UV and optical) are diagnostic of the strength of different regions of the spectrum, and are the primary indicator of ionizing flux. However, as discussed in sections \ref{sec:gasanddust} and shown later in figure \ref{fig:uvlines}, their interpretation is far from simple and multiple line ratios must be used to constrain any specific stellar population. These data can only be interpreted through comparison of data with stellar population synthesis and nebular gas radiative transfer models.

%%%%%%%%%%%%%%%%%%%%%%%%%%%%%%%%%%%%%%%%%%
\section{Stellar Population Synthesis and the Binary Challenge}

%-----------
\subsection{The basics of population synthesis}

A simple stellar population, i.e. that derived from a single molecular cloud collapse, will be parameterised by the shape its initial mass function (i.e. what range of ultraviolet-faint, low mass stars are generated for every ultraviolet-luminous massive star?), the upper mass limit on that function (i.e. is the cloud large enough to generate 50\,M$_\odot$ stars, 100\,M$_\odot$ stars or 300\,M$_\odot$ stars?) and the stellar tracks or isochrones used to generate it. Stellar evolution models (which trace the physical properties of individual stars with time) or isochrones (which trace the properties of a single age population with stellar mass) will depend in turn on their input physics, which may include the effects of non-solar metallicities or compositions, a range of post-main-sequence processes, stellar wind mass loss rates and different prescriptions for mixing and evolution in the stellar interior \citep{2019ses..book.....E}. The entire population is deemed to have a single stellar age which may be young (in which case the most massive, hottest stars dominate) or old (in which case the massive stars have lived and died, and the light will be dominated by the cooler emission of low mass stars).
For applications other than those involving the counting of individual stars or predicting transient event rates, simple stellar population models will usually be enhanced by  coupling  the stellar evolution models to appropriate stellar atmosphere models or empirical spectra. These allow the translation of "hot" and "cold" stellar populations into "blue" or "red" light, and predict more detailed spectral features such as the strength of stellar absorption or emission lines and the shape of the underlying spectral energy distribution.

A composite stellar population is usually more appropriate than simple stellar populations for comparing against unresolved galaxies, since outside of the far-ultraviolet, the light in physical systems is rarely dominated by one single-aged burst of star formation. These composite populations are constructed by weighting together a large number of simple stellar populations according to a stellar cluster mass function, a star formation history or some similar scheme which allows for additional variation in the  distribution of stellar parameters. Now a subset of the stellar population may be old and metal poor, while another subset may be young and metal rich, and each subset can be described by its own mass function. 

The star formation histories considered may be determined in a number of ways. One commonly used analytic approach is to assume that the star formation is decaying exponentially after an initial burst - i.e. SFR $\propto \exp(-t/\tau)$ - or exponential after some initial rise time \citep[e.g.][]{2019A&A...622A.103B}. Alternate approaches might be to assume that the ongoing star formation episode has been constant over a long interval, that it has comprised multiple individual and near-instantaneous bursts with different strengths and ages, or that it can be modelled as a step-wise function with different star formation rates in a predefined number of time bins before the present \citep[e.g.][]{2018MNRAS.480.4379C,2017ApJ...837..170L}. In the optimal case, a star formation history is derived from the data itself through determining the best-fitting linear combination of all possible simple stellar populations, and a posterior probability distribution is determined on this combination \citep{2017MNRAS.472.4297W}. However this is only possible for high signal to noise data, spanning a broad wavelength range, and often a predefined set of template star formation histories is used for fitting. It also adds a large number of free parameters (the weight of each SSP) in the fitting analysis. The complexity of the composite models tested against any given data-set is usually limited by the number of degrees of freedom available in the data rather than the number of ways in which the simple stellar population models can be combined.

\begin{figure}[!ht]
\centering
\includegraphics[width=7.7 cm]{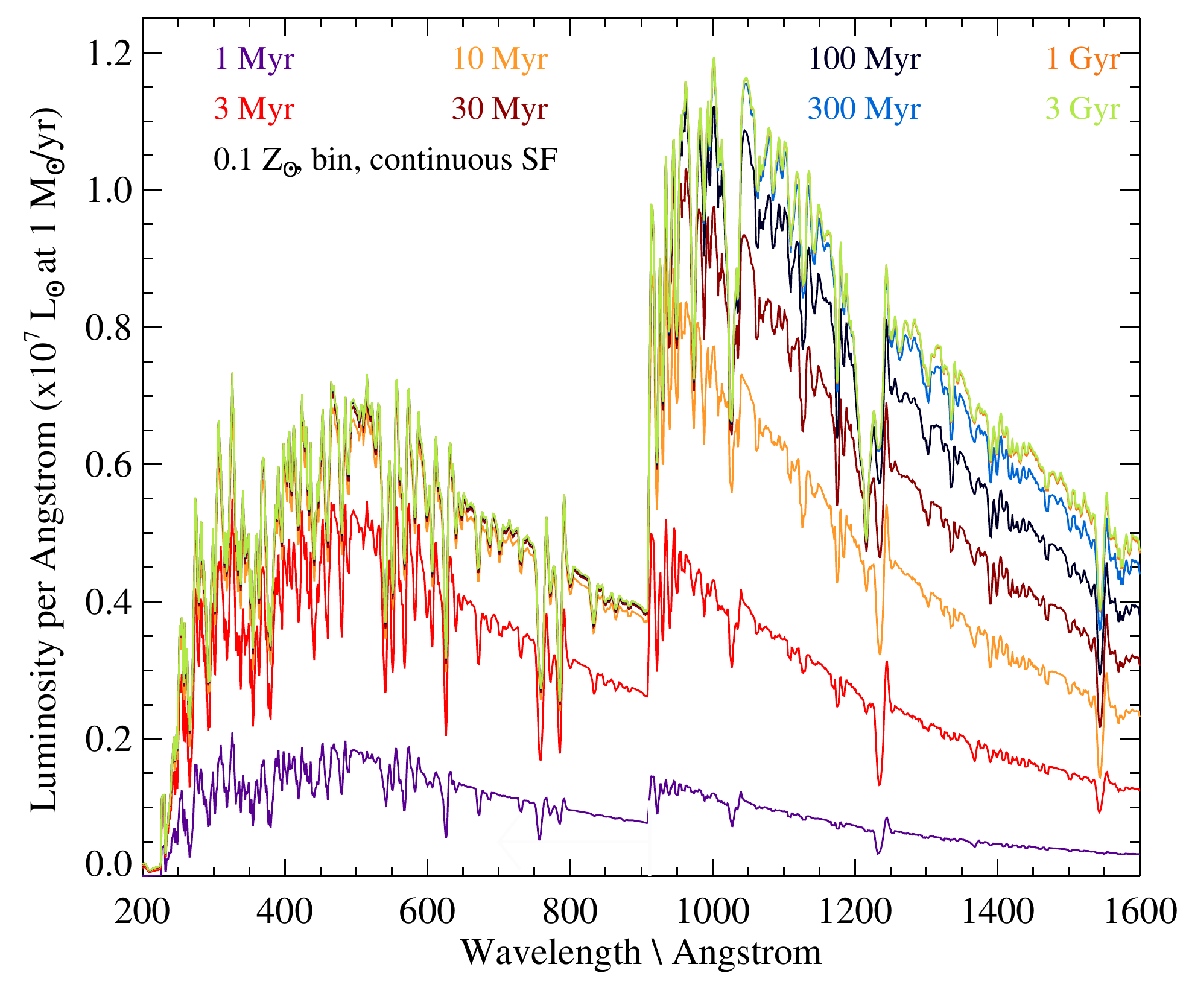}
\includegraphics[width=7.7 cm]{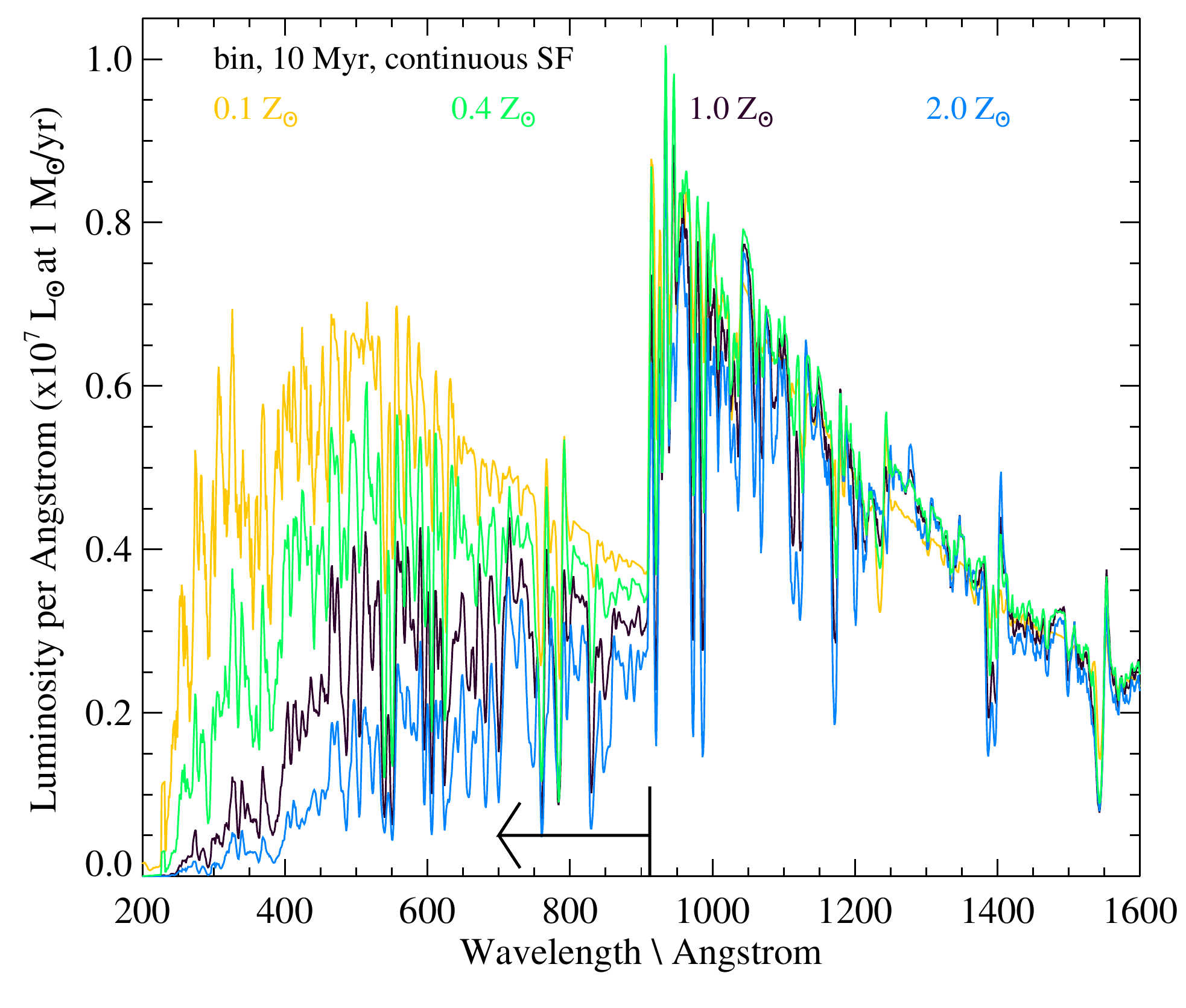}
\includegraphics[width=7.7 cm]{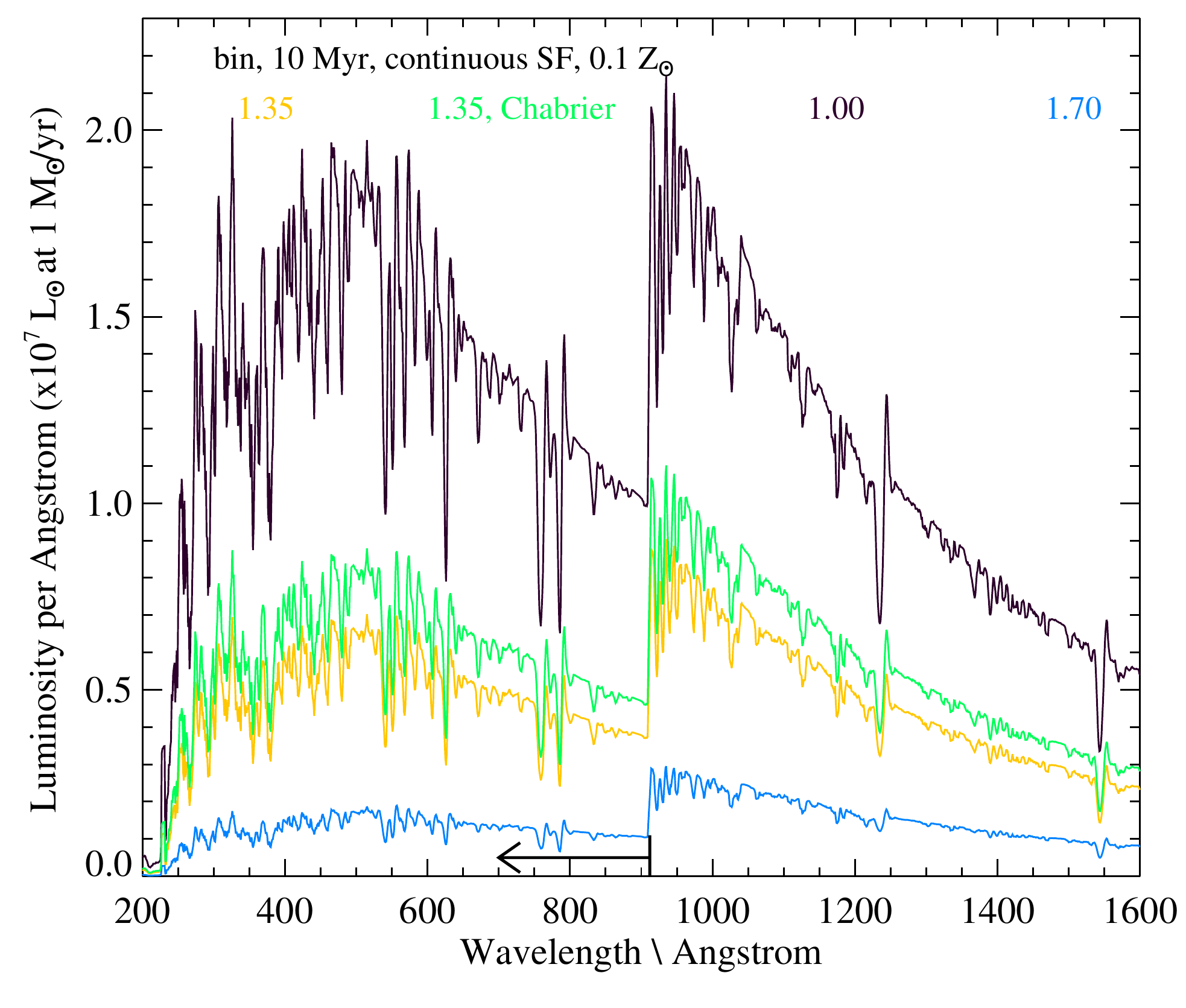}
\includegraphics[width=7.7 cm]{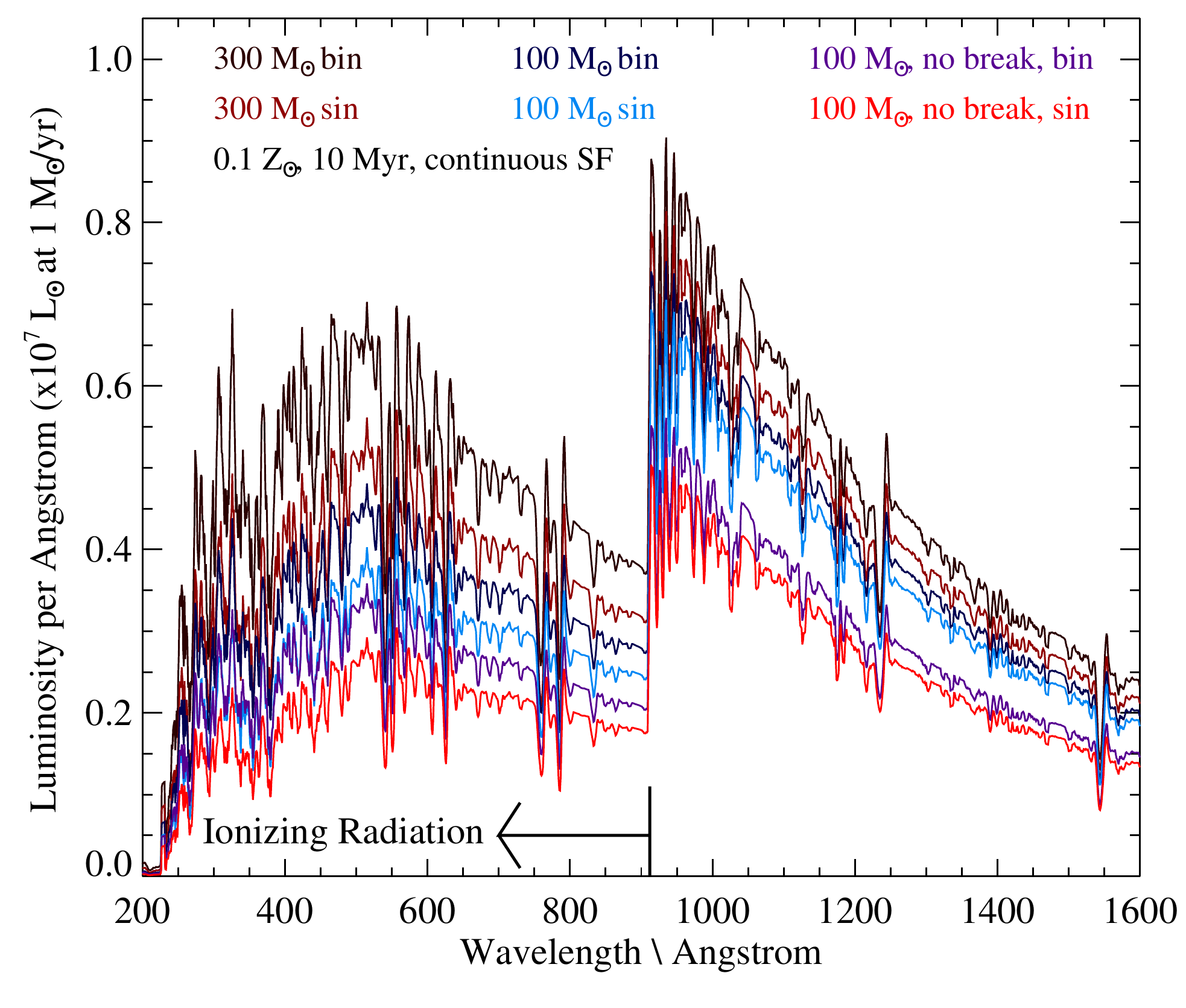}
\caption{The impact of the assumed IMF, age, metallicity and binary fraction on the far-ultraviolet (ionizing) emission from a stellar population produced by BPASS v2.2 \citep{2018MNRAS.479...75S}. Top Left: the effect of age for a population incorporating binary models at Z=0.1\,Z$_\odot$, which has been forming stars continuously at the rate of 1\,M$_\odot$ per year for different lengths of star bursts. The assumed initial mass function is a broken power law with a slope of -1.35 between 0.5 and 300\,M$_\odot$ and -0.3 between 0.1 and 0.5\,M$_\odot$. Top right: the effect on a 10 Myr old binary population with the same IMF of altering the stellar metallicity. Bottom left: the effect on a 10 Myr old binary population at Z=0.1\,Z$_\odot$ of altering the upper slope of the IMF, or replacing the low mass break with an exponential (Chabrier \citep{2003PASP..115..763C}) cut-off. Bottom right: The effect on the same population of altering the upper mass cut-off and of removing the lower break (i.e. a constant IMF slope of -1.35 between 0.1 and 100\,M$_\odot$) on populations incorporating only single or single and binary stellar evolution pathways.\label{fig:seds}}
\end{figure}   
 
\subsection{Improving population synthesis models}

Stellar population and spectral synthesis models have proven extremely powerful in interpreting the light from galaxies in the low redshift ($<z\sim1$) universe \citep[e.g.][]{2003MNRAS.346.1055K,2003MNRAS.344.1000B,2004ApJ...613..898T,2004MNRAS.351.1151B}. In addition to allowing calibration of spectral features such as the Lick absorption line indices as metallicity or age indicators, they have enabled the fitting of SED templates to large photometric samples of millions of galaxies. This procedure yields bulk properties such as the mass, luminosity and redshift of the galaxy, and can distinguish between stellar populations of different ages and even metallicities (although age and metallicity are often degenerate with the degree of dust absorption and its extinction law unless high signal to noise absorption line spectra or a great many photometric bands are available). 

Stellar population synthesis models have also been used to calibrate star formation rate indicators. The assumptions in this process are straightforward \citep{2012ARA&A..50..531K}: 
\begin{enumerate}
    \item Emission in the desired indicator (for example rest-frame ultraviolet 1500\,\AA\ continuum luminosity) is dominated by massive, short lived stars;
    \item For a given flux from massive stars, the total recently-formed stellar mass (which, in contrast to the light, is typically dominated by lower mass stars) can be estimated from the stellar population model;
    \item If the star formation rate has been constant for a long enough period (i.e. over $\sim$100\,Myr), the rate of massive star death balances that of massive star birth and the relationship between flux and star formation rate is linear.
\end{enumerate} 

However, while these methods are widely and robustly applied at low redshift, there are reasons to believe that some of the assumptions on which they are built break down when interpreting galaxies in the more distant Universe and that a new generation of population synthesis models are required. The relevant arguments fall into two main categories:

\textit{\textbf{Metallicity:}}
According to most models of chemical enrichment, the typical metallicity of the interstellar and intergalactic medium in the Universe was some two orders of magnitude lower in the distant Universe than it is today \citep{2014ARA&A..52..415M,2019arXiv191010023P}. Thus the stellar populations that formed from such gas should show evolutionary pathways which differ quantitatively from those seen today. To some extent, this trend is counteracted by observational constraints: the galaxies we observe in the distant Universe are amongst the more massive objects for their redshift, and likely began to form (and so became enriched) earlier than the outlying regions of the cosmic web. However, studies of the mass-metallicity relationship at cosmic noon have shown that star forming galaxies at fixed mass and star formation rate have metallicities lower by $\sim0.1$\,dex at $z\sim2.3$ than at $z=0$ \citep{2018ApJ...858...99S} and this trend is likely to continue to higher redshifts \citep[e.g.][]{2008A&A...488..463M,2014ARA&A..52..415M,2019arXiv191000597C}. Thus relations tested and calibrated in the relatively high metallicity environment of the local galaxy population should not be assumed to hold.

\textit{\textbf{Stellar Age:}}
Massive galaxies in the cosmic twilight of the local Universe are often dominated in mass by old stellar populations ($>$1\,Gyr in age) \citep{2004ApJ...613..898T}. While their light at blue wavelengths may be dominated by star forming regions, these are typically forming stars relatively slowly and at a steady rate over hundreds of millions of years. As a result, interpreting the emission from local galaxies requires a good understanding of long-lived, low mass main sequence stars, together with their evolutionary products and processes such as mass loss on the asymptotic giant branch \citep[e.g.][]{2018MNRAS.479...75S}. By contrast, distant galaxies may emit their light when the Universe itself was less than a billion years old (i.e. at $z>5.7$) and their stars may be forming at rates of tens or hundreds of solar masses per year, in short-lived (tens of Myr) starburst events with rising star formation rates \citep{2007MNRAS.377.1024V}. In models, these processes and their feedback to the environment are highly sensitive to the assumptions made in massive star evolution, and to the upper end of the initial mass function \citep{2017PASA...34...58E}. As a result, distant galaxies probe a different parameter space in population synthesis to the local population.

In this context, it is important to note that local populations do exist which may be used as potential local analogues for the distant galaxy population. These are intensely star forming galaxies, with a stellar metallicity that is typically sub-Solar, and with a comparable specific star formation rate, or star formation rate surface density, to those observed in the distant Universe \citep[e.g.][]{2005ApJ...619L..35H,2014MNRAS.439.2474S}. These galaxies have offered insights into the galactic winds, composition, radio calibration and other properties of such stellar populations, particularly confirming the existence of very hard ionizing radiation fields and an excess of X-ray sources \citep{2016ApJ...822....9H,2017MNRAS.467.4118I,2017MNRAS.471..548I,2017ApJ...847...38Y,2017MNRAS.470..489G,2016MNRAS.459.2591G}. Such populations continue to offer important insights into the properties of star formation in this young, intense, low metallicity regime. However, these galaxies are rare in the local Universe, and their interpretation is often complicated by the presence of older underlying stellar populations, large amounts of dust, and environmental effects not expected to be common in the distant Universe.

The evolution in the properties of typical galaxies towards earlier times, and in particular the shift in importance to massive star evolution, leads to an increased impact from two further properties of the population which are largely overlooked in traditional stellar population synthesis:

\textit{\textbf{Stellar rotation:}} High mass stars are typically born with higher intrinsic rotation speeds than their low mass counterparts \citep{2015A&A...580A..92R}. At low metallicities the stellar wind mass loss rates of stars are lower, and thus they retain more of this initial angular momentum throughout their lifetime, or alternatively retain angular momentum imparted by, for example, mass accretion from a companion \citep{2001A&A...369..574V}. The result is enhanced mixing within the stellar interior, feeding additional hydrogen fuel into the stellar core and resulting in a larger helium core at the end of the main sequence.  Short-lived rotation can  rejuvenate a star, while a star which continues spinning very rapidly through its main sequence lifetime never forms shells burning individual elements but rather undergoes nucleosynthesis in a chemically homogeneous fashion \citep{2016A&A...585A.120S,2016MNRAS.458.2634M}. The result is bluer and hotter stars than might otherwise be expected at a given age.

\textit{\textbf{Stellar multiplicity:}}
The majority of low mass stars are born either alone or in widely-separated binaries that are unlikely to interact on Hubble timescales. By contrast, the multiplicity fraction (i.e. number of stars with binary, tertiary or higher multiple companions) approaches unity for stars with M$_\mathrm{ZAMS}>10$\,M$_\odot$. Indeed the typical number of companions exceeds one. Of these multiples, it has been estimated that 50-70 percent will interact, changing the evolutionary pathway of stars, during their lifetimes \citep{2012Sci...337..444S,2013A&A...550A.107S}. It has further been suggested that the close-binary fraction increases with decreasing metallicity, even for Sun-like stars \citep{2019ApJ...875...61M}, although the evidence for this and the detailed behaviour as a function of mass remains unclear. The result is that the young, massive star dominated, low metallicity stellar populations observed in the distant Universe are expected to host large numbers of interacting stellar binaries. Again, this tends to lead to bluer and hotter stars at a later age than would usually be expected, in part because mass transfer can spin up and rejuvenate the accretor \citep[e.g.][]{2013ApJ...764..166D}, and in part because the interaction extends the lifetime of the more massive star \citep{2014MNRAS.444.3466S}. 

\subsection{The New Generation of Population Synthesis Models}

The availability of massively improved computer processing power, has led to the development of a new generation of stellar population and spectral synthesis models which attempt to address the challenges of modelling stellar populations through a range of approaches.

\subsubsection{Improved Stellar Models}

The field of stellar evolution modelling has developed rapidly in recent years. A range of older, one-dimensional shell-approximation, detailed evolution codes have been supplemented by ever-improving N-body and hydrodynamical simulation codes, and by new  evolution model frameworks, notably the Modules for Experiments in Stellar Astrophysics (MESA) \citep{Paxton2011,Paxton2019}. While studies of individual stars or small systems of stars using such models can now be very detailed, population synthesis requires a large number of models. Not only does the variation in evolution timescale, convective mixing, core burning and stellar radius with mass have to be captured, but also any variants on the behaviour at a single mass.  It may be necessary to consider a range of compositions, a range of stellar rotations or a range of initial binary configurations at any given mass. Hence the grids of stellar evolution models used by population synthesis models must strike a balance between capturing sufficient detail in each model to accurately predict observed populations and retaining sufficient simplicity for the necessary suite of models to be calculable within the available processor time. 

One approach to this is through the use of semi-analytic approximations to the parameter dependence of some or all stages of the stellar evolution, based on interpolation between a sparse existing grid of stellar models \citep[e.g.][]{2002MNRAS.329..897H,2009A&A...508.1359I}. This allows simple estimates of the observed properties to be calculated in fractions of a second rather than minutes. It has the benefit that models can be generated as required, on the fly, rather than requiring a fixed model grid, and also allows  rapid evaluation of the effects of changing initial assumptions regarding the population, or physical parameters such as mass loss rates. As such it is powerful in certain applications such as predicting the rates of astrophysical transients or of specific stellar types in a population \citep[e.g.][]{2008ApJS..174..223B,2018A&A...619A..77K}.

However rapid models will not capture subtleties in the evolution in any situation in which the properties or subsequent evolution of the star is highly sensitive to its interior structure or its state when events such as mass transfer episodes take place. For these, detailed stellar evolution models are required. In particular, the widely-used analytic approximations of \citet[]{2002MNRAS.329..897H} do not consider stars more massive than 50\,M$_\odot$. For several years, 100-150\,M$_\odot$ was estimated to mark the upper end of the stellar initial mass function, based on luminosity-lifetime arguments, assumed stellar wind mass loss rates and the absence of more massive stars from certain clusters \citep{2005Natur.434..192F}. However very massive stars (up to $>$200\,M$_\odot$) have now been demonstrated to exist in the local Universe and show a flattening in the luminosity-mass relation, invalidating such arguments \citep{2018A&A...618A..73S}. Newer stellar evolution models must also generate tracks for these very massive stars \citep[e.g.][]{2017PASA...34...58E}.

Significant advances have also been made in recent years in the simulation of massive stars which undergo rotational mixing during their evolutionary lifetimes. The Geneva grids of stellar evolution models now include a subset in which the star is assumed to evolve while rapidly rotating, leading to additional sheers, meridional circulation, increased internal mixing, turbulence and resultant changes to the interior structure, surface composition and lifetimes of stars \citep{1994A&A...287..803M,2019A&A...627A..24G}. These Geneva models are now implemented in the stellar population synthesis of Starburst99 \citep{2014ApJS..212...14L} and also the FSPS project \citep{2009ApJ...699..486C,2010ApJ...712..833C}.

Meanwhile, models addressing stellar multiplicity have also become feasible with improvements in both modelling and computer capabilities. Significant work developing models for the response of stars to binary mass transfer has been carried out by a variety of teams \citep[e.g.][]{2008ApJ...685..225L,2000A&A...358..462V,2005MNRAS.364..503Z,2017PASA...34...58E,2009MNRAS.400.1019E,2012MNRAS.419..479E}. Very recently, a small grid of atmospheres for stars which have had their envelopes stripped in a binary interaction have been incorporated in a stellar population synthesis by Starburst99 \citep{2019A&A...629A.134G}. However, perhaps the most influential binary synthesis models in recent years have been the products of the Binary Population and Spectral Synthesis (BPASS) project \citep{2009MNRAS.400.1019E,2012MNRAS.419..479E,2016MNRAS.457.1028S,2017PASA...34...58E,2018MNRAS.479...75S}.  These make use of a custom grid of about 250,000 individual detailed, one dimensional stellar evolution models, distributed across 13 metallicities and probing masses from 0.1 to 300\,M$_\odot$, each for a range of binary parameters.

\subsubsection{Improved Atmospheres}

While population synthesis is built on ever-improving stellar evolution models, spectral synthesis also requires developments of improved models for stellar atmospheres.  A stellar atmosphere grid models the spectrum of a star as a function of its effective temperature, surface gravity and surface composition. Luminosity is typically taken as a linear scaling factor during the combination of models, rather than governing the spectrum in itself. A theoretical grid typically builds each model through a radiative transfer calculation including line profile functions and prescriptions for atmospheric turbulence, Doppler and gravity broadening \citep{1979ApJS...40....1K}. Their precision is often limited by the input catalogue of atomic, ionic or molecular species, and the number of transitions considered for each. By contrast an empirical atmosphere grid makes use of observed stellar spectra, fitting each to models in order to estimate their physical properties, and then using either individual spectra or spectral composites as characteristic of stars with those properties \citep{2010MNRAS.404.1639V,1998PASP..110..863P}. The former approach has the benefit of forward modelling: the input physics is fully known and interpretation of matching observed spectra is unambiguous... unless the input physics is wrong.  The latter approach will include all the relevant physics by default, but their interpretation may be uncertain and based on faulty assumptions. For example, empirical spectra may show emission or absorption features that evidently exist in nature, but are omitted from theoretical models because their origin is unknown and cannot be identified. A recent analysis of the effects of library limitations on galaxy spectral fitting was presented by \citet{2019arXiv191011902C}.

The increased complexity of stellar evolution grids in recent years has led to expansions in the parameter space of atmosphere models as well. Not only the atmospheres of main sequence stars and giants need to be included in a spectral synthesis but also the atmospheres of binary products which may, for example, be hotter at a given surface gravity than assumed in standard grids \citep[see][]{2017PASA...34...58E}. One focus of recent attention, particularly in the high redshift universe where stellar populations at $<100$\,Myr are more common, has been improved models for the spectra of stripped, helium-dominated star atmospheres. While either giant Wolf-Rayet stars \citep{2015A&A...579A..75T,2019A&A...621A..85H,2006A&A...457.1015H,2003A&A...410..993H} or naked helium dwarfs \citep{2019arXiv191100543G,2019A&A...629A.134G,2018A&A...615A..78G} may result from extreme wind-driven mass loss in single stars, they result in larger numbers from binary interactions and incorporation of these atmospheres can make significant differences to the ultraviolet spectrum of composite binary stellar populations. To a large extent the production and survival of these extreme objects is dependent on the rate at which stars lose their mass through stellar winds \citep{2018A&A...619A..54V,2018A&A...615A.119V}. These are believed to be weaker at low metallicities than at Solar abundances, although there is still uncertainty in the modelling in the low metallicity regime.

It is also important to note that the ultraviolet spectra (and hence also the ionizing continuum) of stars are poorly understood, particularly at low metallicities ($<$0.1\,Z$_\odot$).  The ultraviolet-absorption of Earth's atmosphere requires observations of local exemplars to be taken from space telescopes. For stars at any significant distance from the Sun, or those which are embedded in nebular gas, the ionizing spectrum is also heavily modified by the intervening medium. As a result, the stellar atmosphere models used for spectral synthesis at very low metallicities are either entirely theoretical or constructed by extrapolation from higher metallicity spectra. However, as discussed later, this lack of empirical calibrators and exemplars for atmosphere models has been recognised by the community and will be addressed over the next few years by the ULYSSES project (see section \ref{sec:unresolved}).

\subsubsection{Key Examples}

An exhaustive list of population and spectral synthesis models is difficult to compile. New models are emerging in the literature, while older models may still be in use in specific applications where their physical inputs remain valid. Here I discuss a few key examples of spectral synthesis models.

The first large evolutionary population synthesis model grid to attract widespread usage in the high redshift community was the \textit{GalaxEv} model suite of  \citet[BC03,][]{2003MNRAS.344.1000B}. This suite has seen revision over the years \citep[e.g. BC03 2016, ][]{2015IAUGA..2257611B}. In particular, the current generation of spectral synthesis models incorporates Wolf-Rayet atmosphere models and MILES stellar models. It now produces a large photon flux at wavelengths below the 912\AA\ break \citep{2019MNRAS.490..978P} relative to the same team's older models. However it remains unclear whether these models, which still include only single stars and thus omit important effects on massive star evolution, provide an accurate description of the distant Universe.

Another stellar population synthesis model which has been applied in the local Universe (e.g. in value-added catalogues for the Sloan Digital Sky Survey), but is not very widely used at high redshift, is the Maraston 2005 model \citep[M05,][]{2005MNRAS.362..799M} and its later modifications \citep{2011MNRAS.418.2785M,2013MNRAS.435.2764M}. Again this is a single star model, but instead of adopting theoretical stellar isochrones it adopts a fuel consumption theorem algorithm to combine the stellar evolution models of  \citet{1997A&A...317..108C}. The resultant outputs are notable for emphasising the role of thermally pulsating asymptotic giant branch stars and the uncertainties introduced by their variability and poorly understood physics in the infrared.

\textit{FSPS} is a relatively new model set which has been built on the MIST stellar isochrone library \citep{2016ApJ...823..102C}. It is unusual in considering a wider range of atmosphere models and stellar compositions than most, and is increasingly being used to constrain the properties of galaxies at $z<2.5$ \citep{2019ApJ...877..140L}.

\textit{MILES} stellar population synthesis models \citep{2010MNRAS.404.1639V} are built on the stellar isochrones of \citet{2000A&AS..141..371G}, combined with a large custom library of empirical stellar spectra, also compiled by the MILES project  \citep{2006MNRAS.371..703S}.

\textit{Starburst99} \citep{1999ApJS..123....3L,2012IAUS..284....2L} is a well-established project generating simple stellar populations (see also review by Leitherer, in this volume).  Its focus has historically been on relatively massive stars and young starbursts, and it was the first large stellar population synthesis code to implement the rotating Geneva stellar models. It has recently incorporated a prescription for including the helium dwarf atmospheres of G\"otberg et al \cite{2019arXiv191100543G,2018A&A...615A..78G} which arise primarily from binary products. 

The \textit{Binary Population and Spectral Synthesis} project (BPASS) \citep{2017PASA...34...58E,2018MNRAS.479...75S}, employs a very large grid of stellar evolution tracks, generated with a custom 1D stellar evolution code which incorporates the effects of interaction and mass transfer between stellar binaries. In addition to an initial mass function, it requires input distributions for initial mass dependence of binary fraction, binary mass ratio and binary period. The effects of rotation are considered in the common envelope phase, and when rejuvenation occurs as the result of binary mass and angular momentum transfer, but stars are typically assumed to rapidly lose acquired spin. A small grid of simple rotating star models (quasi-homogeneous evolution models) is also implemented for massive stars at very low metallicity. Synthetic population outputs include not only spectral data but also transient rates and stellar number counts. The BPASS models are used to illustrate the dramatic effects of IMF and stellar population assumptions on the ionizing photon output of a population in Figure \ref{fig:seds} and combined with nebular radiative transfer models as described below in Figures \ref{fig:uvlines} and \ref{fig:uvcomp}.

Of the models most widely adopted by the extragalactic community, only BPASS and (as of recently) Starburst99 include elements of both rotational and binary effects, although with very different implementations.

\subsection{Nebular Gas and Dust Emission Synthesis}\label{sec:gasanddust}

\begin{figure}[!t]
\centering
\includegraphics[width=7.7 cm]{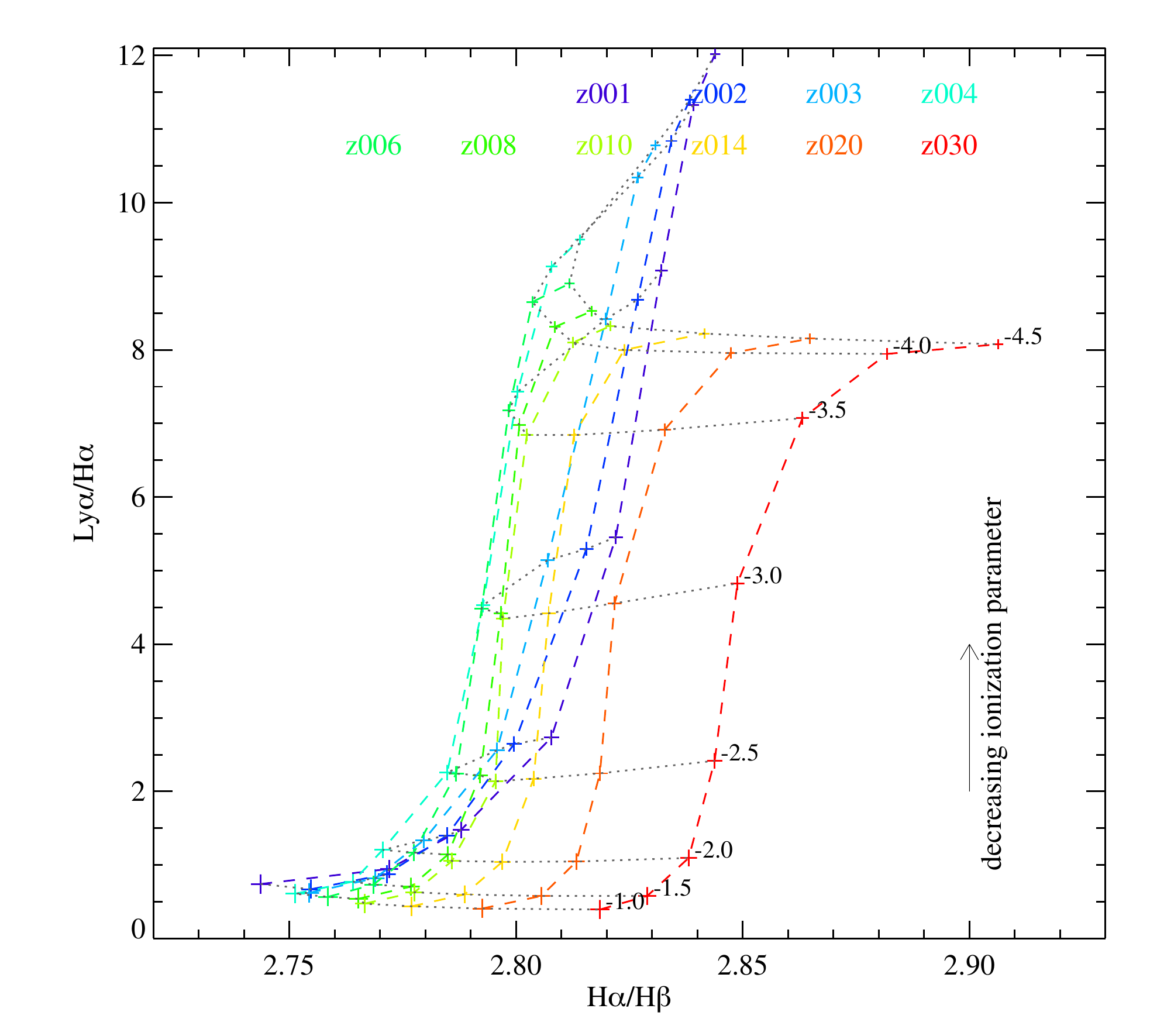}
\includegraphics[width=7.7 cm]{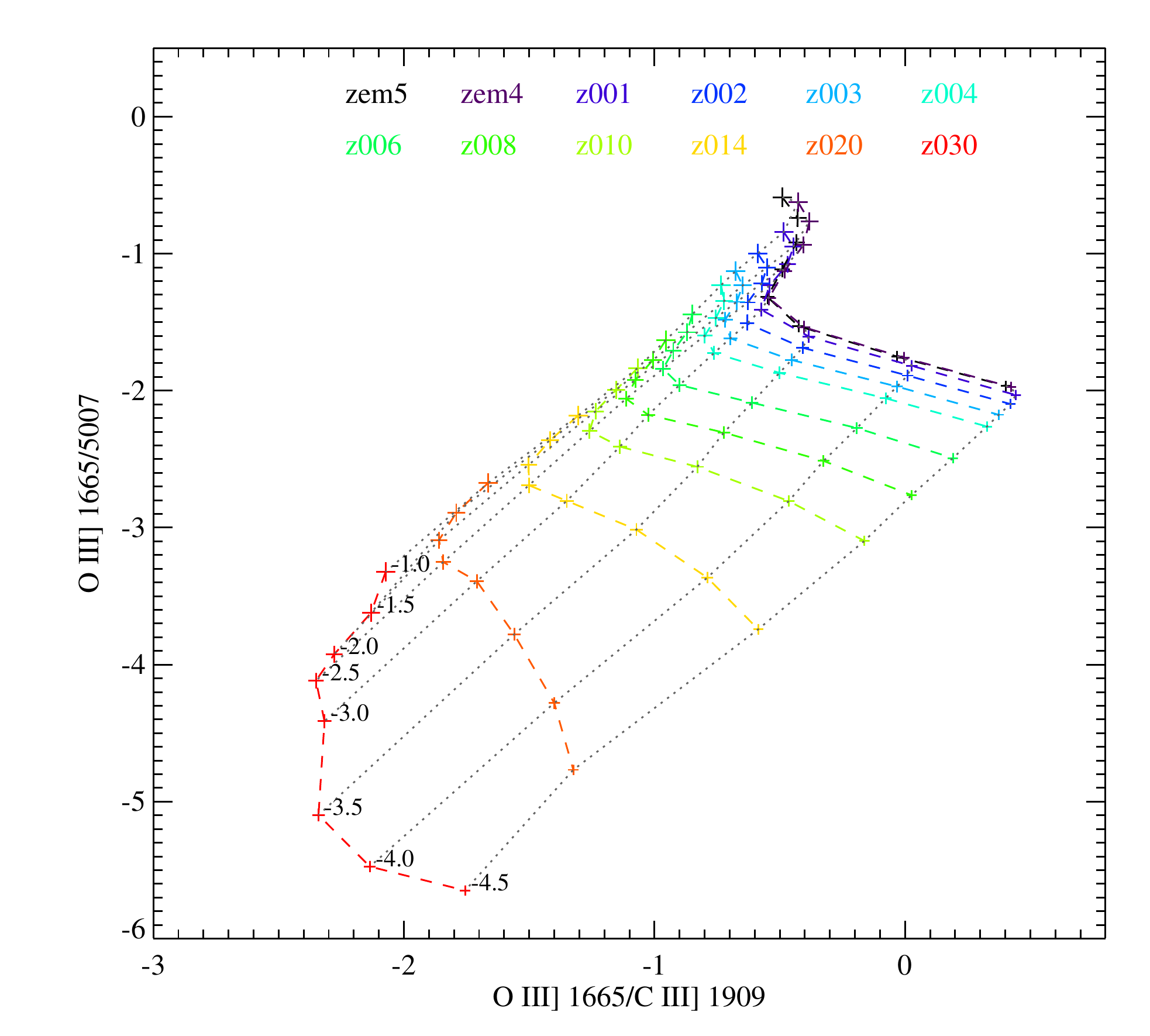}
\includegraphics[width=7.7 cm]{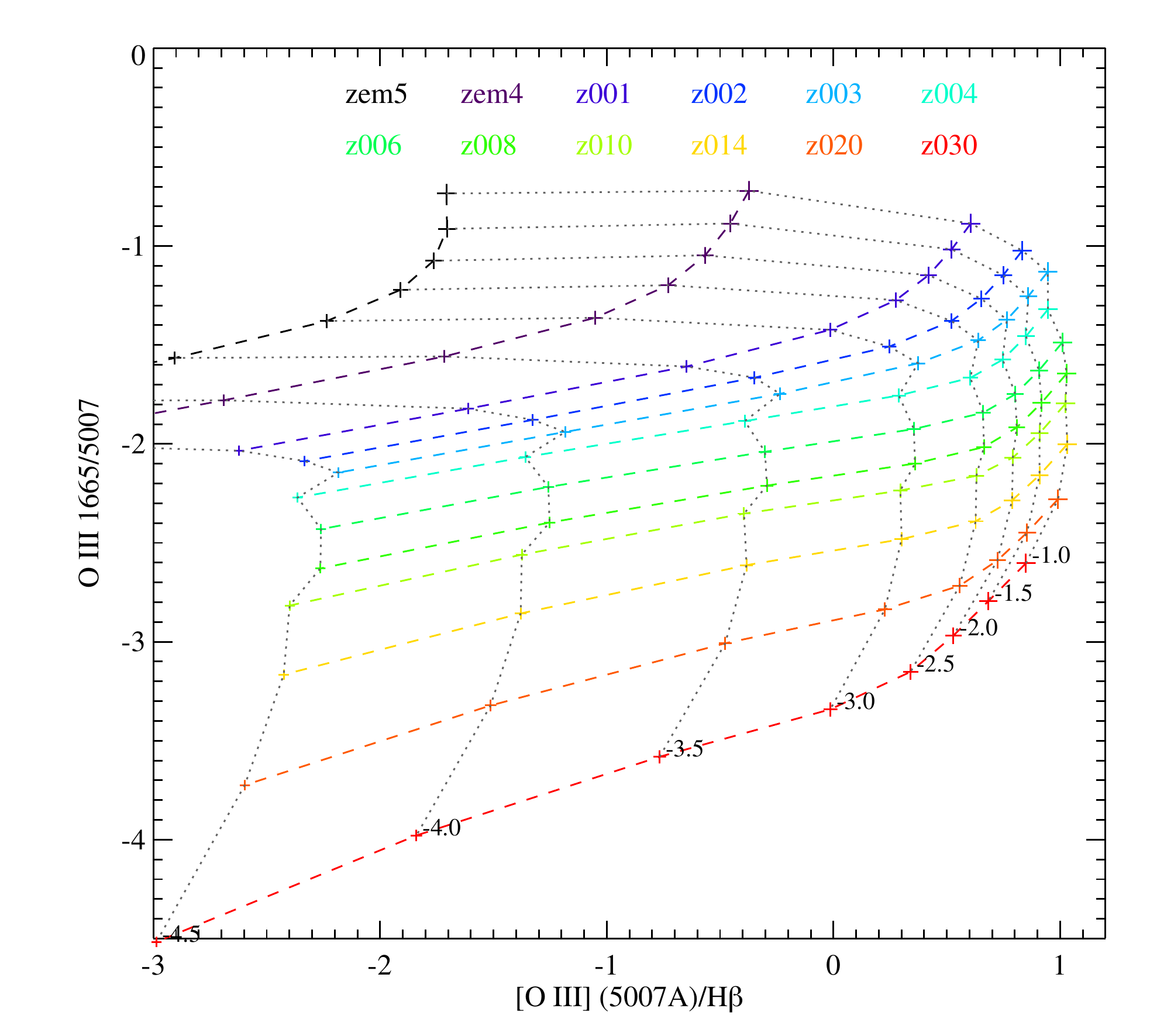}
\includegraphics[width=7.7 cm]{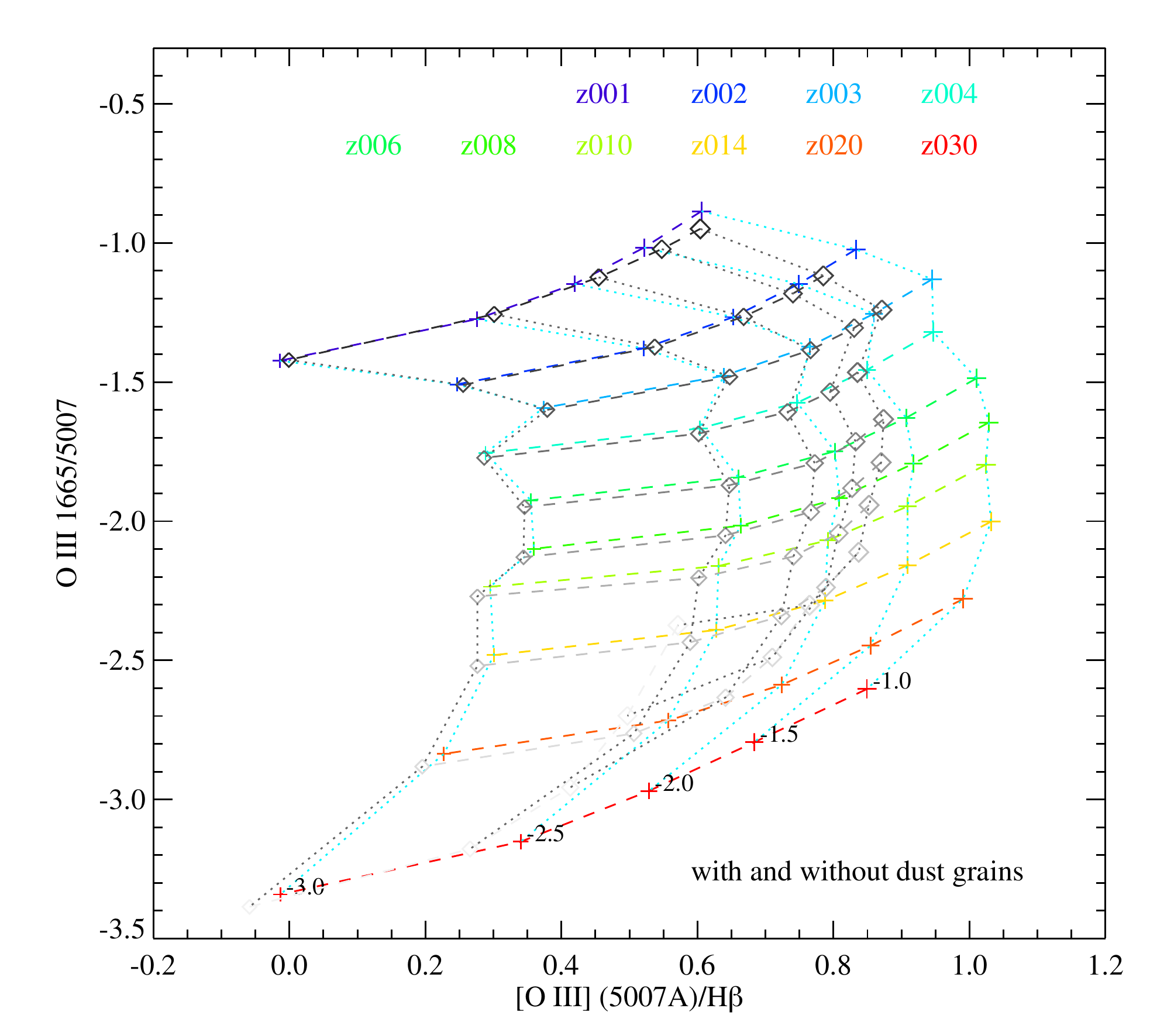}
\caption{The power of ultraviolet line diagnostics to diagnose the physical conditions in ionized gas. In each case the same ionizing spectrum (a continuously star forming stellar population at 100\,Myr) is processed through a nebular gas cloud with the same gas density (log($n_e$/cm$^{-3}$)=2.3) and a range of ionization parameters as labelled on the figures (decreasing this parameter is broadly equivalent to increasing the mean distance of the cloud from the source or decreasing the  star formation rate). In the first three panels we show strong ultraviolet and optical emission line ratios which may be used as ionization and metallicity diagnostics. On the lower right panel we demonstrate the importance of including dust grain surface physics (colour) or omitting it (greyscale) in an otherwise identical radiative transfer model (a subset of ionization parameters and metallicities is shown for clarity). While the effect at low ionization parameters is small, the higher ionization parameters typical of the high star formation densities and binary populations seen in the distant Universe are very sensitive to the assumed physics. All models are from BPASS v2.2 \citep{2018MNRAS.479...75S}, processed with CLOUDY. The metallicity of each line is labelled with `zem4',`zem5', `z001' etc indicating metallicity mass fractions of $Z=10^{-5}, 10^{-4}, 0.001$ etc.\label{fig:uvlines}}
\end{figure}

The light of galaxies is dominated by their stars, but the mass of star-forming galaxies is often dominated instead by their gas content. In the diffuse interstellar medium, this has relatively little effect on the emitted starlight. However newly born stars (i.e. those yet to join the zero-age main sequence) are typically embedded in a region of dense gas and dust. After the onset of core hydrogen burning, the hot young stars, blow out or destroy the dust and ionize the gas to form an H\,II region. The ionized cloud gradually dissipates on timescales of tens of Myr, under the action of radiation pressure and kinetic energy input from stellar winds.  Thus the light from young stellar populations is seldom observed directly, but instead is processed by a screen of nebular gas which can both modify the spectral slope of the continuum (introducing reddening) and re-emit energy in electron transitions of elements (producing emission lines). The majority of energy is emitted in recombination lines of hydrogen (such as H$\alpha$, H$\beta$, Lyman-$\alpha$ etc) and also in metal lines (primarily the forbidden lines of oxygen, carbon, nitrogen and sulphur in the optical).  The relative strengths of these lines depend not only on the gas cloud composition but also its physical properties including density and temperature, and on the spectrum of the ionizing sources \citep[see][and Figure \ref{fig:uvlines}]{2006agna.book.....O}.

The determination of radiative transfer of light through a gas cloud is an entirely different physical process to the calculation of stellar photospheric emission and its evolution. It calls for reprocessing of the spectral synthesis output through specialist radiative transfer codes, prominent amongst which are CLOUDY \citep{1998PASP..110..761F,2017RMxAA..53..385F} and MAPPINGS \citep{1993ApJS...88..253S,2001ApJ...556..121K}. While the spectral synthesis accounts for evolution in the irradiating spectrum, assumptions must be made regarding the gas density and geometry relative to the source (or alternatively the composite of these characterised by an ionization parameter) and also the gas composition, including any interaction between gas particles and dust grains. These are not natural predictions of a stellar population synthesis code and so must be determined empirically, held constant at a given value or estimated from theoretical arguments.

The widely-used BC03 GalexEv population synthesis models \citep{2003MNRAS.344.1000B} applied a nebular gas correction calculated from irradiation by the youngest stellar component. Starburst99 also provides the option to output a nebular emission prescription for a given irradiating spectrum.  Other model sets, including those of BPASS, FSPS and Maraston, have opted instead to increase the transparency of the nebular gas assumptions by releasing the stellar photospheric light alone as core data products. All these teams have also published literature on the gas reprocessing of their models separately, highlighting the uncertainty in the gas assumptions and its potential impacts \citep[e.g.][]{2019MNRAS.482..384X,2018MNRAS.477..904X,2014MNRAS.444.3466S,2017ApJ...840...44B,2019MNRAS.487..333H,2016MNRAS.462.1757G}. While isolating the stellar population spectral synthesis from the gas radiative transfer increases the duplication of effort in undertaking  reprocessing by different members of the community, it also forces each investigator to think carefully about the appropriate assumptions for their science use case.

Nebular gas is not the only interstellar component of galaxies to reprocess starlight. A further obstacle between emitted stellar photospheric photons and the observer can be found in the form of dust grains. These absorb light preferentially at blue wavelength, according to an extinction law which depends on grain size and composition. The photon energy heats the grains to temperatures of tens of Kelvin, and is reemitted in a form that can be approximated as thermal blackbody radiation (with the precise spectrum again potentially modified by grain properties) which peaks in the far infrared \citep{2000ApJ...539..718C,2002ApJ...576..159D}. 
Again, this process must be accounted for before comparison between any given stellar population synthesis code and observational data, particularly in galaxies with sufficient dust extinction to modify their spectral shape. In the optical this may require imposing a wavelength-dependent suppression of the model flux, the strength of which must be determined through emission line ratios where available and through fitting a range of extinctions otherwise. In the infrared and longer wavelengths, the dust emission component becomes important, and in the mid-infrared yet another cooling mechanism must be considered: the excitement of broad emission lines in large polyaromatic hydrocarbon molecules.

The MAGPHYS  \citep{2008MNRAS.388.1595D,2015ApJ...806..110D},  CIGALE \citep{2002A&A...393...33B,2019A&A...622A.103B}  and PROSPECTOR models \citep{2017ApJ...837..170L}  are all SED generation and fitting codes which provide algorithms which self-consistently balance the re-emission to the absorbed energy for a given dust model, allowing fitting of the full SED from ultraviolet to radio wavelengths against observational data. These algorithms take stellar population synthesis models as their input (currently FSPS for PROSPECTOR, BC03 for MAGPHYS and BC03 or M05 for CIGALE).

\begin{figure}[!th]
\centering
\includegraphics[width=7.7 cm]{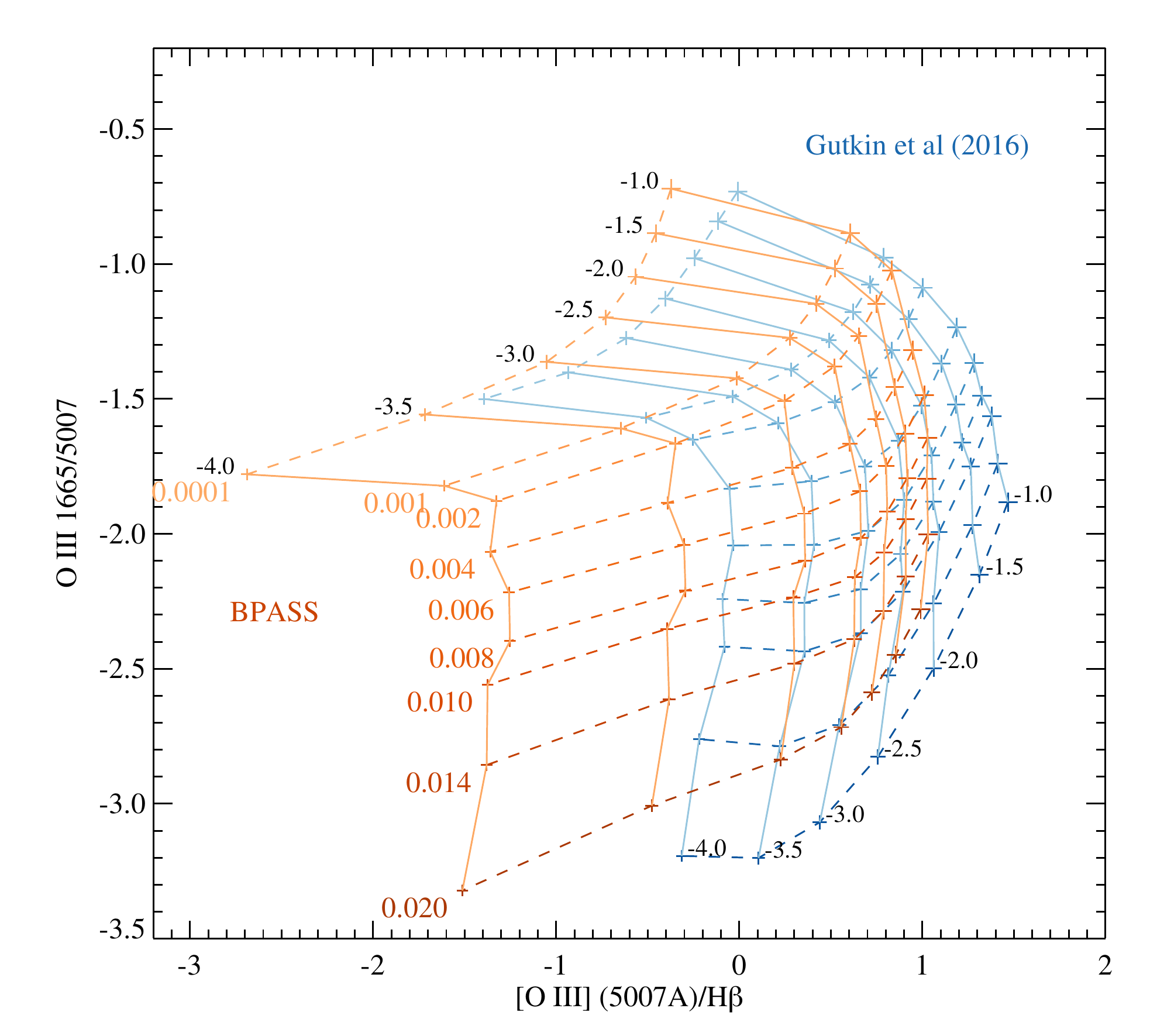}
\includegraphics[width=7.7 cm]{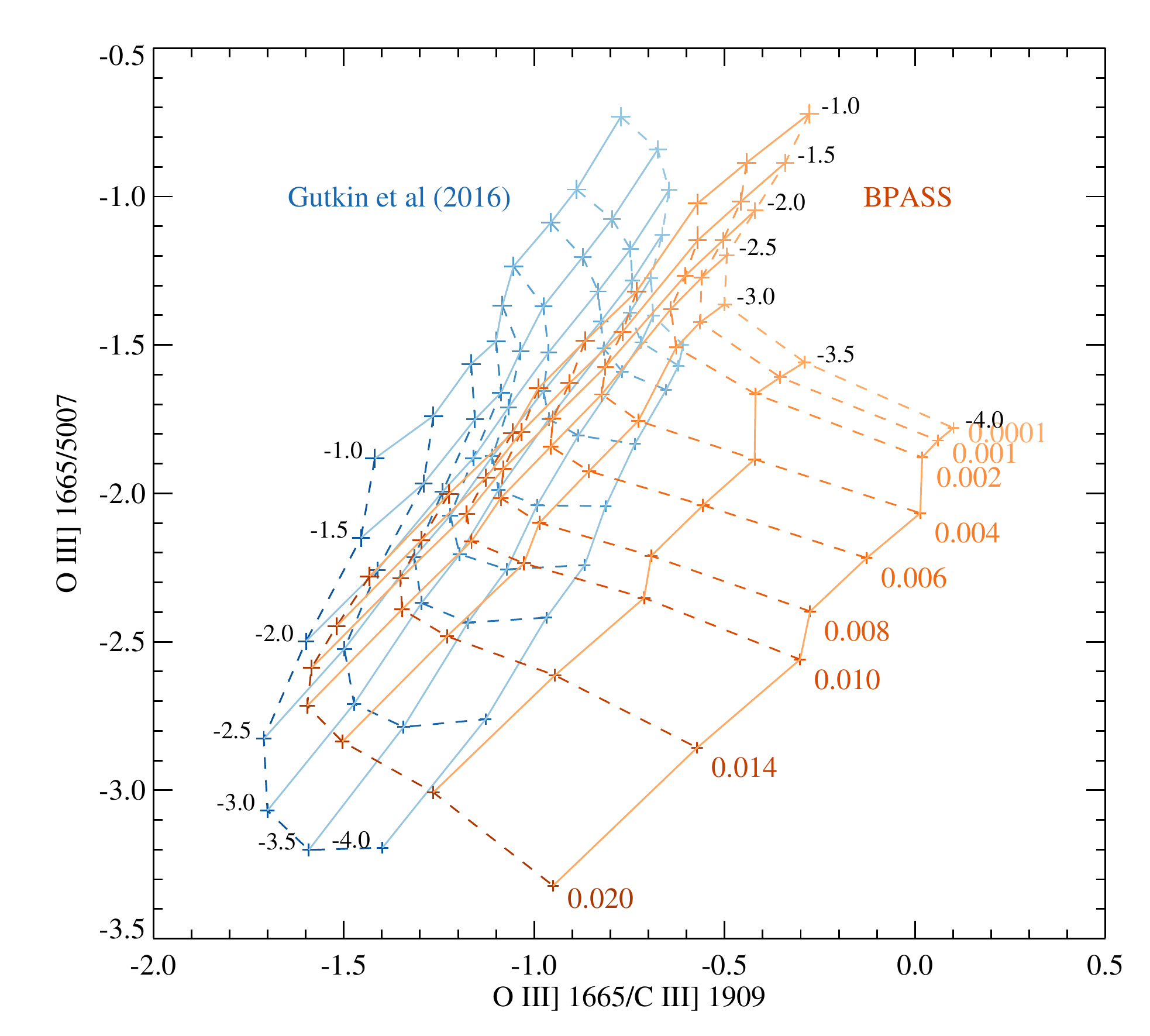}
\caption{The effect of these stellar population choice on these diagnostic diagrams. Models compared are the BPASS models as before (Orange) and the nebular synthesis models of \citet{2016MNRAS.462.1757G} (Blue, based on the 2016 version of the GalaxEv models of BC03). BPASS models assume a broken power law IMF, with an upper slope of -1.35 and an upper mass cut-off at 300\,M$_\odot$, while the nebular gas electron density is 300\,cm$^{-3}$. The \citet{2016MNRAS.462.1757G} model assumes an IMF with the same slope and upper mass limit, but a \citet{2003PASP..115..763C} lower mass cut-off, for $n_e=100$\,cm$^{-3}$, [C/O]=1 and a dust-to-metal ratio of 0.5. Both models assume constant star formation and are shown at matched metallicities. The flux in the ultraviolet doublets has been summed before ratios are taken. The primary difference is in the handling of stellar evolution and atmosphere. \label{fig:uvcomp}}
\end{figure}   
 
%All figures and tables should be cited in the main text as Figure 1, Table 1, etc.

%\begin{table}[H]
%\caption{This is a table caption. Tables should be placed in the main text near to the first time they are cited.}
%\centering
%%% \tablesize{} %% You can specify the fontsize here, e.g., \tablesize{\footnotesize}. If commented out \small will be used.
%\begin{tabular}{ccc}
%\toprule
%\textbf{Title 1}	& \textbf{Title 2}	& \textbf{Title 3}\\
%\midrule
%entry 1		& data			& data\\
%entry 2		& data			& data\\
%\bottomrule
%\end{tabular}
%\end{table}

%%%%%%%%%%%%%%%%%%%%%%%%%%%%%%%%%%%%%%%%%%
\section{Model-Critical Science Results}

The examples of stellar population synthesis models applied in the distant Universe are too numerous to discuss in detail. Many have already been discussed above. Here we highlight key examples of prominent science results and discussion in the literature where the interpretation is heavily dependent on the assumed stellar population synthesis model.

\begin{enumerate}
    \item The epoch of reionization marks the transition between the neutral intergalactic medium of the Dark Ages to the highly ionized IGM that followed the epoch of Cosmic Dawn. The detection of this transition, through Lyman-$\alpha$ line damping and the identification of polarised scattering in the cosmic microwave background radiation, is unambiguous. However a number of details regarding its sources and interpretation are highly model dependent
    
    \begin{enumerate}
        \item As has already been discussed, and as is shown by Figure \ref{fig:seds}, the ionizing photon production rate of a given population is highly sensitive to its modelled parameters, amongst them the stellar binary fraction and stellar population age. Figure \ref{fig:seds} is presented for the continuous star formation case, which in theory minimises the differences between populations. However it is still clear that the adoption of different assumptions, including the timescale of the star formation epoch, its metallicity and IMF can cause large differences in the ionizing flux. In particular, small variations in stellar metallicity can dramatically change the ionizing photon production without producing large changes to the ultraviolet continuum longwards of the Lyman break (Figure \ref{fig:seds}, upper right panel). Thus the model assumed can have a large effect on discussions of whether the observed galaxy population produces sufficient photons to drive reionization, on the role of the faint end slope of the luminosity function and the minimum stellar masses required for high redshift galaxies \citep[see e.g.][]{2016MNRAS.458L...6W}.  As Figure \ref{fig:uvlines} (upper left) indicates, Balmer line strengths and their ratios depend only weakly on the nebular conditions, and an indirect estimator of the ionizing photon production efficiency can be derived from the H$\alpha$ line flux powered by this radiation. However a number of authors have taken this analysis a step further, first inferring the H$\alpha$ strength from Lyman-$\alpha$ emission. Since the resonant Lyman-$\alpha$ line is highly dependent on the nebular conditions (see Figure \ref{fig:uvlines}), ambiguous interpretation is inevitable.\smallskip

        \item The escape fraction of ionizing photons also plays a crucial role in the above calculations. If photons are produced in stellar photospheres but cannot escape their galaxy haloes, their impact on the IGM will be negligible. There are two major impediments to photon escape: line-of-sight dust extinction and nebular gas absorption. A common assumption in past work has been that the dust extinction of ionizing flux shortwards of 912\AA\ directly tracks that of the observed 1500\AA\ continuum, such that the photons observed in the latter can be used to infer the former. However doing so requires the application of a dust extinction law and there are large uncertainties in the shape of the extinction curve at short wavelengths, which is determined by the dust grain size and composition but which cannot be directly measured \citep{2017MNRAS.470.3006C,2015ApJ...806..259R,2011A&A...532A..45S}. This introduces large uncertainties in the ionizing flux escape due to the dust correction, even before gas extinction is considered.  
        
        Estimates of the escape fraction in galaxies can be obtained in a number of ways, but perhaps the most important are direct imaging below the Lyman limit or  analysis of the nebular emission lines, in both cases comparing to the inferred photon flux expected from stellar population models. As already discussed, nebular emission estimates can be heavily affected by assumed nebular gas conditions (Figure \ref{fig:uvlines}) and by the shape of the assumed irradiating spectrum, as shown in Figure \ref{fig:uvcomp} which compares matched models in the BPASS model grid and those derived by \citet{2016MNRAS.462.1757G} from the BC03 2016 release. As  \citet{2016MNRAS.456..485S} demonstrated, the tension between the high escape fractions required by reionization modellers and the low typical escape fractions seen in galaxies can be reconciled if models producing a higher photon flux (such as binary models or those at slightly lower metallicity) are used. \smallskip

        \item A related challenge is determining which physical processes affect the escape fraction and whether it changes over time - i.e. are galaxies with high fractions sparse in gas and dust, or simply patchy, with  channels of near unity escape fraction existing in an otherwise opaque medium? In the  `picket-fence' scenario, the likely culprits for blowing such channels out are core-collapse supernovae, but for short-lived starbursts, the ionizing flux in most models has subsided before supernovae begin to occur. By contrast, as \citet{2016MNRAS.459.3614M} and \citet{2018MNRAS.479..994R} have demonstrated, the longer ionizing lifetimes of stars in models including interacting binary star populations make them a prime candidate for the sources of the escaping photons which power reionization. A similar effect might be obtained from repeated bursts of star formation on short timescales.

    \end{enumerate}

\item
The cosmic volume-averaged star formation density evolution is inferred from the luminosity function of Lyman break galaxies and their conversion into star formation rate functions through model-calibrated factors. As already discussed, these factors are heavily dependent on the assumptions made for stellar populations \citep[see e.g.][]{2019MNRAS.tmp.2490W}. Equally important, but perhaps less obvious at first glance, is the dependence of the rest-UV inferred star formation history on corrections for dust extinction. Since these are inferred from fitting the observed colours of a galaxy (or occasionally its emission line ratios or spectrum) with a template and a dust curve, there is inevitable dependence on the stellar population synthesis model used. The bluer spectra predicted  by models dominated by steep or top-heavy IMFs, and by binary or rotating stars, may suggest that the dust correction may be larger than previously estimated using models which exclude these effects.

\item Interpretation of the stellar populations responsible for stellar wind driven feedback, outflows, enrichment of the intergalactic medium and thus the cosmic chemical evolution all rely on interpreting the spectra of distant sources, or on making assumptions based on their luminosities, masses and star formation rates in a limited handful of mostly rest-ultraviolet bands. For example modelling chemical enrichment from a stellar population requires assumptions for their star formation rate (model-dependent),  core collapse supernovae time delays and rate distributions (model dependent \citep{2019MNRAS.482..870E}) and white-dwarf or compact merger rates (model dependent \citep{2016MNRAS.462.3302E,2019MNRAS.482..870E}).  In fact, as has been extensively discussed above, and as demonstrated by Figures \ref{fig:uvlines} and \ref{fig:uvcomp}, all such interpretations rely on the models being used. Indeed many of the simple models may well fall short of being physical in most scenarios: while observers routinely compare the strengths of oxygen and carbon line features (as in Figure \ref{fig:uvcomp}, for example) the C/O ratio is itself time-dependant, with different production timescales for the two elements. No simple stellar population synthesis model currently incorporates the details of this evolution in elemental abundance ratios. While this may seem to be less true of evolutionary properties inferred from large scale cosmological simulations (i.e. N-body or hydrodynamic models) of dark matter halo growth, the galaxy evolution models which are built on the haloes are in turn informed and validated against model-dependent interpretations of observational data.

\end{enumerate}

%%%%%%%%%%%%%%%%%%%%%%%%%%%%%%%%%%%%%%%%%%
\section{The Holistic View}

As has been emphasised in previous sections, many stellar population and spectral synthesis models are optimised for specific applications. This permits limited resources (whether computational or in the form of investigator time) to be channelled into specific issues which affect that application, while the modelling of other issues may be presumed to follow comparatively simple prescriptions. For example a binary model intended to study merging compact objects may focus on the physics of the critical common envelope phase, rather than the main sequence and giant branch evolution of the component stars. This approach has been implicit in most applications to galaxy populations until recently: if the majority of galaxies to be studied are giant systems, a Hubble time in age, with their surviving stellar populations dominated by Solar metallicity, Solar mass or lower stars, the binary fraction of massive stars and their interaction physics may legitimately be considered a low priority for resource allocation. 

However, optimizing models for one application carries a risk: models may be fined tuned to match a small subset of observational data, while neglecting evidence from other sources that suggest their solutions are not universally applicable.
We can take a simplistic but nonetheless striking example. Spectra for young galaxies may be fit with either single star population models or those including binaries. The single star models have been fine tuned and demonstrated to perform well in fitting typical (i.e. old, metal-rich) galaxies in the local Universe. In fitting the young galaxies, the derived parameters will differ between model grids and the quality of the fits may be comparable. Since the single star model has fewer parameters (not requiring initial binary distributions, common envelope parameters and other physics) it may well provide the formal best fit. Occam's razor suggests the model with fewest parameters, in this case the single star model, should be accepted.  However there is a problem with this: the binary fraction for high mass stars in the local Universe is unity, and that for Solar mass stars is around 20 percent. Thus unless something is radically different with the physics of star formation in the galaxies under consideration (itself a significant assumption), there must be binary stars in these systems. Therefore the single star models are not the correct models. They are in fact models in which an important prior assumption has been made: that the binary fraction, mass ratio distribution and period distribution are all zero at all primary masses. Overlooking this strong - and unsupported - prior can lead to misinterpretation of the posterior probability distributions on the derived parameters. 

As processing power continues to increase, becoming less of a constraint in some situations than available researcher time, a new approach to validating stellar population synthesis modelling has become possible. The key premise of this approach is that a detailed understanding of any aspect of stellar population modelling is likely to be inaccurate, and may even be entirely incorrect, unless all available constraints are also fully accounted for. 

This approach is not trivial to implement, and any such implementation will necessarily be incomplete. Nonetheless, this is the philosophy adopted by the BPASS stellar population synthesis project. Rather than releasing purely simple stellar population spectral models, and their derivatives (e.g. synthetic photometry, ionizing photon rates and Lick spectrophotometric indices), the BPASS project also releases expected stellar type number counts, stellar remnant parameter distributions, nucleosynthetic yields for chemical evolution models and the rates of a range of electromagnetic and gravitational wave transients. Crucially these are not calculated in a separate synthesis, but directly represent the population for which the spectral data is derived. This allows the models to be checked on multiple levels: not only are individual stellar models benchmarked against observational data, but so are the properties of both resolved  and unresolved stellar populations, and the rates of different transient types as a function of both cosmic time and metal enrichment. The current generation of models does, of course, fail in some aspects. In particular, the strength of very high ionization emission lines such as He\,II\,1640\AA\ in distant galaxies continues to present a challenge for every stellar population synthesis code, BPASS included \citep[see][and discussion therein]{2019A&A...621A.105S}. However with a wide range of constraints considered, it is possible to check whether changes to the population synthesis which attempt to address this issue remain consistent with other observations.

An example of the multiple-test approach may perhaps be seen in the combination of recent papers which use BPASS.  \citet{2019MNRAS.482..870E} used the supernova rate output of the stellar population synthesis to demonstrate that population synthesis models which incorporate massive binaries can simultaneously recover the cosmic evolution in the observed volumetric rate of core collapse transients and the estimated volumetric rate of compact binary mergers as probed by short Gamma-Ray Bursts and Gravitational Wave transients.  At the same time \citet{2019MNRAS.tmp.2490W} have demonstrated that the same stellar population models, when combined with an appropriate spectral synthesis, can improve the agreement between the \textit{in situ} stellar mass density and the volume-averaged star formation rate density. These models have not been fine tuned or rescaled to fit either observational data set, and while neither fit is exactly perfect, both are comparable to the quality obtained by other models. Thus the ability of the model to fit two highly-independent observational constraints provides a reassuring validation of the physical assumptions embodied within it.  Obviously, as with any model, there are areas which might be improved, in this case the dust and metallicity properties of the galaxies could be further constrained. Nonetheless, with each additional constraint the model is further validated or areas for future work identified.

%%%%%%%%%%%%%%%%%%%%%%%%%%%%%%%%%%%%%%%%%%
\section{Unresolved Challenges}\label{sec:unresolved}

A great many comparisons between stellar population synthesis models have been performed over the years, each identifying a mixture of strengths and weaknesses \citep[see e.g.][]{2018MNRAS.476.4459D,2010A&A...515A.101C,2010ApJ...712..833C}. Enormous progress has also continued to be made in recent years, particularly as regards consideration of stellar multiplicity effects. Nonetheless there remain unresolved challenges in the field of stellar population and spectral synthesis, particularly where it is applied to galaxies in the distant Universe.

 Stellar mass is perhaps the most robustly recovered property of galaxies from stellar population synthesis modelling, although a variation of $\sim0.3$\,dex between models is not uncommon. The mass estimate is essentially provided by the normalisation of the spectral energy distribution in the (rest-frame) optical, and the mass to light ratio of galaxies in this wavelength regime varies only slightly over a range of population models \citep[see][]{2018MNRAS.479...75S}. Binary star models tend to predict slightly higher masses for young stellar populatons. In most scenarios, incorporation of rotating or binary stars in a synthesis shifts the spectral energy distribution towards bluer wavelengths and increases the mass to light ratio, as does increasing the upper mass limit on the initial mass function. The result has been a reevaluation of the cosmic volume-averaged stellar mass density, which has been demonstrated to improve the agreement between this and the integrated star formation rate density history, when both are evaluated using the same stellar population models \citep{2019MNRAS.tmp.2490W}. 

It is also now possible to obtain largely robust estimates for the relative excitation states of nebular gas in star forming galaxies at high redshift, although the choice of emission lines for this analysis needs to be considered carefully. Many of the strongest emission lines in both the rest-frame ultraviolet and rest-frame optical are forbidden lines, which depend on the electron density and temperature of the nebular gas \citep[see][]{2006agna.book.....O}. Where ratios are used between different element species (i.e. [O\,III]/H$\beta$) the effects not only of the irradiation spectrum and subsequent excitation, but also of enhanced abundance ratios between the elements must also be considered. In particular, the enrichment of the interstellar medium by supernovae (on timescales of a few Myr to tens of Myr) in the distant Universe is expected to lead to significant enhancement of $\alpha$-process elements \citep[e.g.][]{2008ApJ...685...40W,2013ApJ...774...64W,2018MNRAS.478.1795H}. Studies of galaxies at $z\sim2-3$ have suggested that the iron-to-oxygen abundance ratios as inferred from Fe-dominated stellar continuum and O-dominated nebular emission features can only be reconciled if such an $\alpha$-enhancement is assumed \citep{2016ApJ...826..159S}.  In this context, it should be noted that  most of the strong emission line ratios used as abundance indicators \citep{2008A&A...488..463M,2019ARA&A..57..511K} are in actual fact excitation measures. Their calibration is based on the principle that lower metallicity stellar populations produce harder ionizing spectra. As has been seen, this is a model-dependant assumption and will be vulnerable to any evolution in the initial mass function and binary parameters with metallicity or redshift, as well as uncertainties in the typical stellar population age or star formation history. Spectroscopy with the NIRSPEC instrument on the soon-to-launch James Webb space telescope will enable these studies to be extended to higher redshifts, and to fainter galaxies near Cosmic Noon. The ability to combine sensitive spectroscopy on targets from the rest-frame ultraviolet through to the rest-frame optical enables the stellar population to be determined with more fidelity than studying either wavelength range alone. While relative calibration of the (observed) optical and infrared can be challenging, achieving this opens the capacity to probe multiple lines from the same species and break some of the degeneracies between ionizing spectrum and abundance ratios. An example of a potentially accessible line ratio - that of the O\,III]\,1665\AA\ and [O\,III]\,5007\AA\ lines - which may serve to distinguish metallicity from ionization parameter is shown in Figure \ref{fig:uvcomp}. Since this line ratio is also one which shows significant differences between different population synthesis model grids, it may also serve to evaluate the differences between these and constrain parameters such as the interacting binary fraction.

The initial mass function, rotation distribution, binary fraction, mass ratio and period distribution all remain large uncertainties in stellar population synthesis modelling.  While the local binary parameters are now reasonably well constrained \citep{2017ApJS..230...15M}, it is essentially unknown in any other environment. Even the initial mass function remains difficult both conceptually and in terms of establishing its properties and universality \citep{2018PASA...35...39H}. The properties of the upper end of the mass function are also unclear. While some of the very massive stars (M$>100$\,M$_\odot$) in the Large Magellanic Cloud are clearly single stars, it is possible that they represent very rapid, early mergers within $<1$\,Myr of reaching the main sequence. If this is the case, then perhaps a $\sim$150\,M$_\odot$ Zero Age Main Sequence limit is indeed appropriate, with 200\,M$_\odot$ stars modelled only as merger products, although whether this is consistent with the number densities of these objects remains unclear \citep{2018A&A...618A..73S}. Since the inferred initial mass function in the LMC is also highly overabundant in very massive stars compared to the canonical Salpeter IMF \citep{2018A&A...618A..73S}, the treatment of this upper limit can have dramatic consequences. One constraining factor on the stellar upper mass limit may well be the dependence of stellar wind strength on metallicity. The sparse extant observations of low metallicity massive stars \citep[e.g.][]{2000PASP..112.1243W,2014A&A...572A..36T,2017IAUS..329..313G,2019arXiv190804687G} do not agree on whether the mass-loss rates scale with previous assumptions for the metallicity \citep{1995ApJS...99..173L,2000A&A...362..295V}. 

The dependence of multiplicity, rotation and other stellar properties with metallicity, particularly very low metallicites,  remains largely an open question and observations of galaxies in the distant Universe may indeed help to constrain these by excluding inappropriate combinations of these model parameters. The current best estimates for multiple fraction, binary period and binary mass ratio as a function of primary mass are currently derived at near Solar metallicity \citep{2017ApJS..230...15M}. These are consistent with a binary (or higher) fraction of one at masses about 16\,M$_\odot$, and these are biased towards close binaries, which will interact. The best estimate of multiple fraction at intermediate masses (5-9\,M$_\odot$) is 76$\pm$8\% and this considerable range of uncertainty theoretically permits binary interactions to extend ionizing photon production toward later times (a higher binary fraction towards lower masses) or to shorten the ionizing lifetimes (a lower close binary fraction). Some constraints on this can be derived from the populations of stripped envelope supernovae or interacting neutron star binaries, although other uncertainties \citep[for example on the residual surface hydrogen fraction permitted for stripped supernovae, or the surface stellar wind strength, see e.g.][]{2019MNRAS.tmp.2830C} may be degenerate with those on the binary fraction itself.

There are, of course, also regimes accessible in the local Universe which remain to be investigated further. Key amongst these, particularly for the interpretation of ionizing spectra from distant stellar populations, is the prospect of improving our understanding of stellar atmospheres in the ultraviolet. The {\em Hubble Space Telescope} Director's Discretionary programme "UV Legacy Library of Young Stars as Essential Standards" (ULLYSES) is building a uniform library of massive standard star spectra in the ultraviolet, including those at a range of metallicities, as well as obtaining spectra for a range of lower mass stars. This programme emerged as a result of consultation on the optimal legacy from {\em Hubble's} ultraviolet capabilities and has been approved to obtain 1000 orbits of data from Cycle 28 onwards. These empirical spectra, inevitably, will not span the full range of parameter space probed by stellar population synthesis models, particularly at the lowest metallicities currently being modelled, but will certainly inform the physics implemented in atmosphere modelling and provide key calibration points. Such an initiative is urgently needed, given the wide range of ultraviolet predictions derived from different stellar libraries and stellar population synthesis models.

Other local insights may also be possible. For example, arguably the power of Atacama Large Millimetre Array (ALMA) to measure the excitation, composition and physical conditions of systems in the mid- to far-infrared has been under-exploited to date as a constraint on synthesis models. This is largely since emission in the ALMA bands is dominated by the reprocessing mechanisms discussed in Section \ref{sec:gasanddust} above. Nonetheless, with the James Webb Space Telescope soon to revolutionise our understanding of the near to mid-infrared in the nearby Universe, as well as on the optical wavelengths at high redshift, such constraints on stellar populations  may well prove important for further development in this field in future.

%%%%%%%%%%%%%%%%%%%%%%%%%%%%%%%%%%%%%%%%%%
\section{Conclusions}

The use of stellar population synthesis models to interpret star forming galaxies in the ultraviolet and at longer wavelengths is widespread and an area of ongoing development. In particular, observations of galaxies in the distant Universe present a particular challenge, not least because it is impossible to resolve their stars and double check the calibration of stellar population synthesis models in regimes poorly sampled nearby. Nonetheless, progress has been made on a number of fronts to improve the modelling of the young, low metallicity and perhaps binary-rich populations expected in these sources. As additional, complementary probes of early star formation become available - perhaps in the form of core-collapse transient rates, massive binary merger rates, or rest-frame optical spectroscopy of high redshift galaxies from space - these models will continue to improve, shedding new light on the stellar populations of the earliest galaxies.

%%%%%%%%%%%%%%%%%%%%%%%%%%%%%%%%%%%%%%%%%%
\vspace{6pt} 

%%%%%%%%%%%%%%%%%%%%%%%%%%%%%%%%%%%%%%%%%%
%% optional
%\supplementary{The following are available online at \linksupplementary{s1}, Figure S1: title, Table S1: title, Video S1: title.}

%%%%%%%%%%%%%%%%%%%%%%%%%%%%%%%%%%%%%%%%%
%\authorcontributions{For research articles with several authors, a short paragraph specifying their individual contributions must be provided. The following statements should be used ``conceptualization, X.X. and Y.Y.; methodology, X.X.; software, X.X.; validation, X.X., Y.Y. and Z.Z.; formal analysis, X.X.; investigation, X.X.; resources, X.X.; data curation, X.X.; writing--original draft preparation, X.X.; writing--review and editing, X.X.; visualization, X.X.; supervision, X.X.; project administration, X.X.; funding acquisition, Y.Y.'', please turn to the  \href{http://img.mdpi.org/data/contributor-role-instruction.pdf}{CRediT taxonomy} for the term explanation. Authorship must be limited to those who have contributed substantially to the work reported.}

%%%%%%%%%%%%%%%%%%%%%%%%%%%%%%%%%%%%%%%%%%
\funding{This research was funded in part by United Kingdom Science and Technology Facilities Council (STFC) grant number ST/P000495/1.}

%%%%%%%%%%%%%%%%%%%%%%%%%%%%%%%%%%%%%%%%%%
\acknowledgments{The author acknowledges her ongoing and productive collaboration with Dr J J Eldridge (University of Auckland) on the BPASS project, the development of which underlies much of the work discussed here.}

%%%%%%%%%%%%%%%%%%%%%%%%%%%%%%%%%%%%%%%%%%
\conflictsofinterest{The author declares no conflict of interest.} 

%%%%%%%%%%%%%%%%%%%%%%%%%%%%%%%%%%%%%%%%%%%
%%% optional
%\abbreviations{The following abbreviations are used in this manuscript:\\
%
%\noindent 
%\begin{tabular}{@{}ll}
%MDPI & Multidisciplinary Digital Publishing Institute\\
%DOAJ & Directory of open access journals\\
%TLA & Three letter acronym\\
%LD & linear dichroism
%\end{tabular}}

%%%%%%%%%%%%%%%%%%%%%%%%%%%%%%%%%%%%%%%%%%%
%%% optional
%\appendixtitles{no} %Leave argument "no" if all appendix headings stay EMPTY (then no dot is printed after "Appendix A"). If the appendix sections contain a heading then change the argument to "yes".
%\appendix
%\section{}
%\unskip
%\subsection{}
%The appendix is an optional section that can contain details and data supplemental to the main text. For example, explanations of experimental details that would disrupt the flow of the main text, but nonetheless remain crucial to understanding and reproducing the research shown; figures of replicates for experiments of which representative data is shown in the main text can be added here if brief, or as Supplementary data. Mathematical proofs of results not central to the paper can be added as an appendix.
%
%\section{}
%All appendix sections must be cited in the main text. In the appendixes, Figures, Tables, etc. should be labeled starting with `A', e.g., Figure A1, Figure A2, etc. 

%%%%%%%%%%%%%%%%%%%%%%%%%%%%%%%%%%%%%%%%%%
\reftitle{References}

\end{document}